\documentstyle[12pt,aaspp4]{article}
\newcommand{\PSbox}[3]{\mbox{\rule{0in}{#3}\includegraphics{#1}\hspace{#2}}}
\newcommand\mathnew{\mathsurround=0pt}
\def\simov#1#2{\lower .5pt\vbox{\baselineskip0pt \lineskip-.5pt
        \ialign{$\mathnew#1\hfil##\hfil$\crcr#2\crcr\sim\crcr}}}
\def\msun{$M_{\odot}$}
\def\epsimin{$\epsilon_{\rm min}$}
\def\epsiminn{$\epsilon^0_{\rm min}$}

\lefthead{Wen L.}
\righthead{Kozai effect and gravitational wave detection}

\slugcomment{\bf Submitted to ApJ.}

\begin{document}

 \begin{center}
\title{On the Eccentricity Distribution of  Coalescing Black Hole
Binaries Driven by the Kozai Mechanism in Globular Clusters}
\author{Linqing  Wen\altaffilmark{1}}
\affil{Division of Physics, Mathematics, and Astronomy\\ California Institute of Technology}
\altaffiltext{1}{Address: Caltech LIGO, MS 18-34, 1200 E. California Blvd., Pasadena, CA 91106. Emails:  lwen@ligo.caltech.edu}
\authoremail{ lwen@ligo.caltech.edu}
\today
\end{center}

\begin{abstract}
In a globular cluster, hierarchical triple black hole systems
can be produced through binary-binary interaction.  It has been
proposed recently that the Kozai mechanism could drive the inner
binary of the triple system to merge before it is interrupted by
interactions with other
field stars.   We investigate qualitatively  and numerically the
evolution of the eccentricities in these binaries under gravitational
radiation (GR) reaction. We predict that $\sim 30$\% of the
systems will possess  eccentricities $>0.1$ when their  emitted
gravitational waves pass through 10 Hz frequency.  The implications
for gravitational wave detection, especially the  relevance to data
analyses for broad-band laser-interferometer gravitational wave detectors, are discussed.
\end{abstract}
\keywords{gravitational waves---binaries: close---stellar dynamics---relativity}

\section{Introduction}
\label{intro}
Globular clusters are excellent birthplaces for black hole (BH) 
binaries.   In our Galaxy, a globular cluster (GC)  contains typically $N=10^6$
stars and resides normally in the galactic 
halo.    Typical ages of these clusters
are estimated to be 12 billion years, and the clusters contain the oldest
stars known in our Galaxy.  Stellar evolution theory
predicts that massive stars with
mass above 10 \msun\ would evolve into black holes or neutron stars  in $\sim 10^7$ years. For
a typical GC that follows
the initial mass function (IMF) given by Scalo (1986), a significant
number of BHs ($\sim 6\times 10^{-4} N$ for initial masses $> 20$--25\msun) should have been born.  These
BHs would be the most 
massive stars in the cluster, and hence dynamical two-body 
relaxation causes them to  sink rapidly towards
the core.   In the core, BH binaries are expected to form.  The single BHs form 
binaries preferentially with another BH, and BHs born in binaries
with a lower mass stellar component rapidly exchange the companion
 for
another BH through three body relaxation processes.  If these BH binaries merge within the Hubble
time, they would become excellent sources for upcoming gravitational
wave (GW) detectors (e.g. \cite{thorne87}). 

It has been argued that globular clusters are unlikely places 
to  harbor BH mergers (or subsequently to harbor a massive central
black hole) if we consider binary-single 
interactions alone (e.g. \cite{simon00}).   Numerical simulations show
that binary-single interactions tend to throw out these BH binaries
and only 8\% of BHs will be retained in the GC
within its lifetime.   Miller \& Hamilton (2002) argued that this
picture will be different if we consider binary-binary
interactions.  A substantial fraction of
hierarchical triple BH systems can result from binary-binary interactions.  The
 well-known Kozai mechanism (\cite{kozai62}) can then drive the inner binaries of a
significant 
fraction of these triples to extreme eccentricities and thus reduce their
time to coalescence to be shorter than the Hubble time.  This is
because the coalescence time
scale $\tau_{\rm GR}$ due to  gravitational radiation  for  a binary
with fixed semi-major axis $a$  has a strong dependence on the eccentricity
$e$ through the quantity $\epsilon=1-e^2$  (\cite{peters64}), 
\begin{equation}
\tau_{\rm GR} \approx 6\times 10^{10}\frac{(a/{\rm AU})^4
(\epsilon/0.01)^{3.5}}{(m_0+m_1)m_0m_1/\mbox{\msun}^3}\ \ \ \mbox{yr},
\label{tau_GR}
\end{equation}
where $m_0$ and $m_1$ are the masses of each binary component.  In their paper, Miller \& Hamilton (2002)
have compared the coalescence time scale with the time scale for the system to be
disrupted by interactions with field stars. They concluded that
in a substantial fraction of cases, the system will achieve extreme
eccentricities ($\epsilon
\sim 0$)  and 
will merge by gravitational radiation 
before the disruption.  The BH mergers
may thus be retained in the GCs and a central intermediate-mass BH may form through
subsequent mergers.

We are motivated by the fact that the proposed mechanism promises
 stellar mass BH-BH mergers with extremely high initial
eccentricities and that they are potential sources for upcoming
 gravitational wave detectors such as LIGO. We investigate in this paper whether
 or not these systems would still possess finite eccentricities when
 the emitted gravitational waves  enter LIGO's frequency
 band.   The eccentricities of these sources are
 particularly interesting as no 
 astrophysical merger sources were expected before to possess
 significant eccentricities when they reach the LIGO band.  Current
 efforts in detecting GWs from coalescing BH  binary sources
 thus focus on waves from circular orbits (see \cite{Ali02} for recent
 development). On the other hand,  
it is well-known that the gravitational waveform of an
eccentric binary could be significantly different from that of a
circular one and different detection strategies are required for
optimal detections (e.g., \cite{peters63}; \cite{peters64};
\cite{lincoln90}; \cite{moreno95}; \cite{gopa02}; \cite{glam02}). 
 Our work can thus help 
 shed  light on the importance of the eccentricities for detecting GWs from these sources.

In this paper,  we investigate in detail the
eccentricity evolution of  these merger systems under the impact of
 gravitational radiation reaction (simplified as the GR effect). This is in addition to the Kozai
mechanism and the post-Newtonian periastron precession (simplified as
the PN effect) considered by previous authors.  It has been long realized (\cite{peters63};
\cite{peters64}) that gravitational radiation reaction is very
efficient in circularizing coalescing binaries.   For the systems we are considering, this circularization effect would
have to compete with the ``eccentricity enhancing'' Kozai
mechanism. Under the Kozai mechanism and the PN effect, the system
tries to maintain secular cyclic
eccentricity oscillations at fixed amplitude (with exceptions at
critical inclination angles, see text).  The GR
effect operates most efficiently near the  minimum $\epsilon$ of each
cycle. In some cases, it  could drive a system near its minimum
$\epsilon$ directly towards merger.  In other cases, it would affect the
minimum $\epsilon$ and alter the otherwise fairly strict periodic
evolution trajectory 
of the system. In particular, it will enhance rapidly the importance of the
PN effect which in turn would introduce fast oscillations to destroy the Kozai resonance.  The Kozai resonance
will eventually be quenched either directly by the GR effect, or by the PN
effect before the GR
effect takes over. We will discuss the details of these effects and their implications for the evolution of the eccentricities and gravitational wave frequencies.

In section~\ref{kozai}, we give a summary of the relevant features of
the Kozai
mechanism and derive an explicit  general expression for the minimum $\epsilon$ (or maximum eccentricity)
the system can reach for arbitrary values of BH masses and initial
eccentricities. In section~\ref{GRR}, we discuss the evolution of 
the system parameters under gravitational radiation reaction (the GR effect).  The
 equations of motion are described in
subsection~\ref{GP}.  We describe technical details to  estimate the minimum
$\epsilon$ the
system can reach in subsection~\ref{epsm}.  The PN effect
and its role in determining the  evolution of the gravitational wave 
frequency are described in subsection~\ref{PN}.  A procedure to
obtain the relation of the gravitational wave frequencies and the
eccentricity at various evolution stages is given in
subsection~\ref{s_f_GW}. Two numerical examples are presented in
subsection~\ref{example}.   In section~\ref{eccn}, we estimate the eccentricity distribution when the gravitational waves enters the LIGO band. Our results are then summarized in section~\ref{concl}.

\section{Kozai Mechanism}
\label{kozai}
\subsection{Principle and Notation}
The Kozai Mechanism was initially proposed (\cite{kozai62})  to
describe the perturbation of asteroids due to Jupiter.  This
mechanism describes the secular evolution of the orbit of a hierarchical
triple system  as a result of cyclic exchange of angular momentum
between the inner binary and orbit of the distant tertiary.  One of the most interesting
results of the Kozai mechanism is that the eccentricity of the inner
binary can reach very large values while the semi-major axis remains
unchanged,  provided that the initial mutual
inclination angle of the two binaries reaches a  certain critical
value. Although the mechanism was proposed for a system where the mass
of one component in the inner binary is much smaller than those of the
others,  the mechanism has then been found to operate in  hierarchical
triple systems with arbitrary masses (\cite{lidov76}).

\begin{figure}
\PSbox{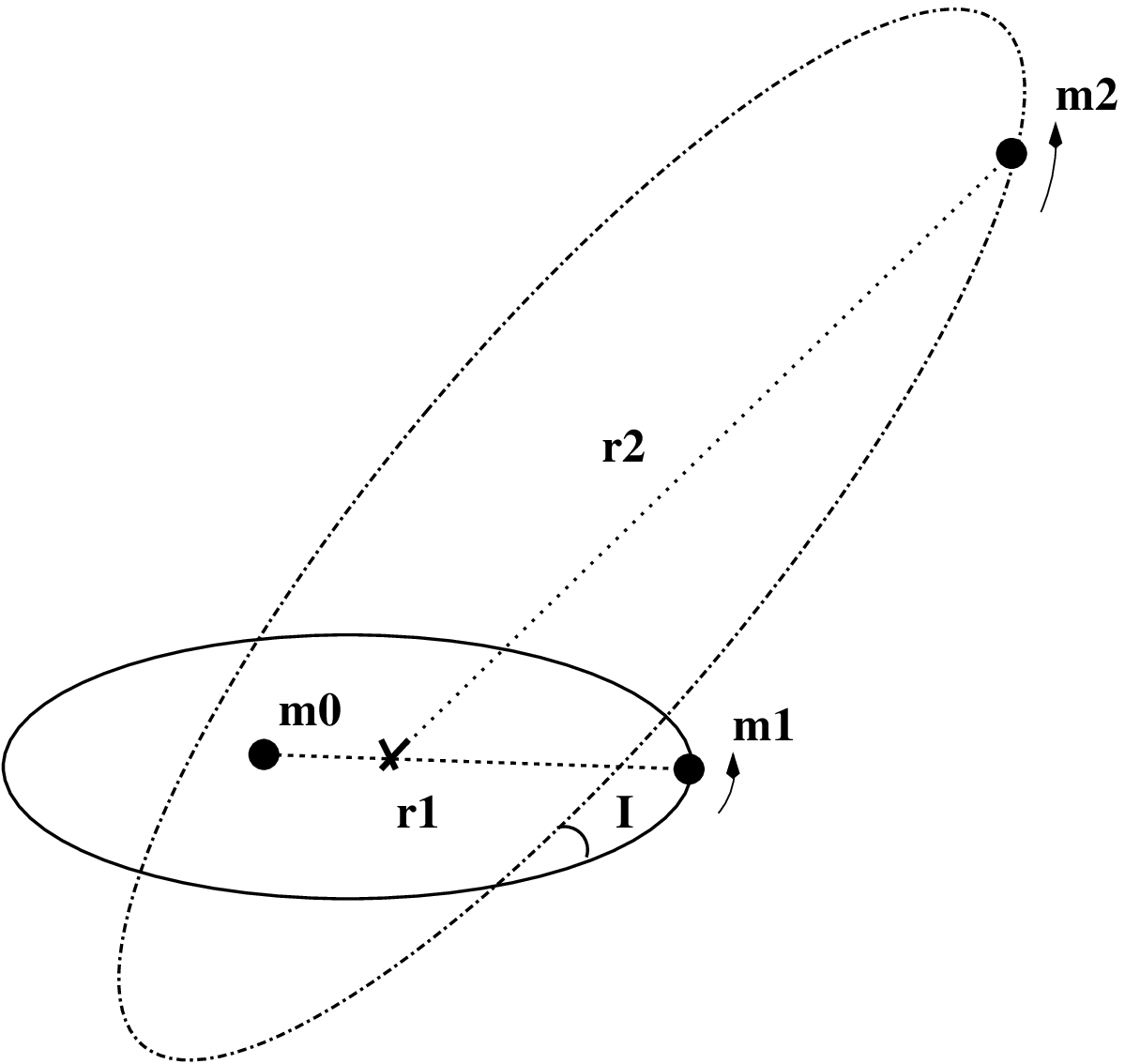 hoffset=70 voffset=70
hscale=90 vscale=100}{6.5in}{6.5in}
\caption{Geometry of a hierarchical triple system.}
\label{geo}
\end{figure}

A hierarchical triple star system consists of a close binary of masses
$m_0$, $m_1$,
and a tertiary $m_2$ in a much wider
outer orbit (Fig.~\ref{geo}).  The dynamics of the system can be
treated as if it consists of an inner binary of point masses $m_0$ and
$m_1$, and an outer binary of point masses $m_2$ and
$M_1=m_0+m_1$.  The separations of the masses each binary are
denoted $r_1$ and $r_2$.  The mutual inclination angle of
the two binaries is denoted $I$.  For the
inner and outer binaries respectively, we denote the eccentricities $e_1$ and $e_2$,
 reduced masses $\mu_1=m_0m_1/M_1,
\mu_2=m_2M_1/(M_1+m_2)$,  semimajor axes  $a_1$ and $a_2$. We define
$g_1$ as the argument of pericenter of the inner binary relative to
the line of the descending node.  

Throughout this paper, we frequently use quantities $\epsilon=1-e^2_1$ and
$b_2=a_2\sqrt{1-e^2_2}$ to simplify our discussion.  Angular momentum
is normalized to have dimension $\sqrt{\mbox{Length}}$ by dividing by the quantity
$L_1=\mu_1\sqrt{GM_1}$. We use notation $\alpha_0$, $\beta_0$, and $\gamma$ for the normalized
magnitude of total angular momentum  of the triple system, of the outer
binary, and of the inner binary respectively. Our $\alpha_0$ and
$\beta_0$ are related to 
dimensionless quantities $\alpha$ and $\beta$  used in previous
literature by 
$\alpha=\alpha_0 /\sqrt{a_1}$, and $\beta =\beta_0/\sqrt{a_1}$.  It follows that
$\gamma=\sqrt{a_1\epsilon}$,
$\beta_0=\mu_2/\mu_1\sqrt{M_2a_2/{M_1}(1-e^2_2)}$, and that $\alpha_0$
 is related to $\beta_0$ and $a_1\epsilon$ through the following
trigonometric relation
\begin{equation}
\cos I=\frac{\alpha^2_0-\beta^2_0-a_1\epsilon}{2\beta_0\sqrt{a_1\epsilon}}.
\label{cosI}
\end{equation}
As we shall see, during the binary's Kozai mechanism-induced evolution,
$\alpha_0$, $\beta_0$, and $a_1$ are conserved while $\epsilon$
changes.  It follows that $\alpha_0 =
\beta_0$ is the critical condition for the system
to be able to evolve to  $\epsilon = 0$ ($e_1=1$).  This implies  a
critical initial $I$ of
\begin{equation}
I_0=I_c =\cos^{-1}\left (-\frac{\sqrt{\epsilon_0}}{2\beta}\right ),
\label{Ic}
\end{equation}
where $\epsilon_0$ is the initial $\epsilon$.  For this solution to exist, $\beta \ge \sqrt{\epsilon_0}/2$ is
required.

The total Hamiltonian of the system can be written as
$H=H_1+H_2+H^{'}$, where $H_1$ and $H_2$ are the Hamiltonians for the 
inner and outer binaries respectively as if they were isolated
binaries of point masses. $H^{'}$ is the perturbative Hamiltonian
which includes perturbation of the inner binary by the presence of the
tertiary $m_2$  and perturbation of the outer binary by the
finite size of the binary component $M_1$.   For $r_1/r_2 \ll 1$ and
$(m_2/M_1)(r_1/r_2)^3 \ll 1$, the Hamiltonian $H^{'}$ can be expanded in
powers of $r_1/r_2$. The first non-zero contribution comes from the
term of order 
$(r_1/r_2)^2$, labeled as quadrupole term because of the degree of
the Legendre polynomial associated with it.  This term is sufficient
for our discussion as we are 
only concerned about cases with large initial mutual inclination angle
$I_0$ (see \cite{ford00} for discussions concerning the octupole terms
of order $(r_1/r_2)^3$).

It is this perturbative Hamiltonian $H^{'}$ that leads to a secular evolution
of the triple system. In the first approximation, the secular
evolution can be described by $H^{'}$ 
doubly-averaged over the mean anomaly of both the inner and the outer
orbits.  It turns out that the doubly-averaged $H^{'}$, the total angular momentum of the
system ($\alpha_0$),  the energy of each
binary (or equivalently $a_1$ and
$a_2$), and the magnitude of the angular momentum of
the outer binary ($\beta_0$, or equivalently $e_2$) are conserved quantities.  The
secular evolution of the triple system can therefore be fully
described in terms of time-evolving  $\epsilon$ and $g_1$.  A  useful constant
of integration can be derived from equation~(\ref{cosI}),
\begin{equation}
a_1\epsilon \left (\cos I+\frac{\sqrt{\epsilon}}{2\beta}\right
)^2=\frac{(\alpha^2_0-\beta^2_0)^2}{4\beta^2_0}.
\label{aecosI}
\end{equation}
Note that the quantity on the left hand side is conserved  as long as the total angular momentum ($\alpha_0$)  and that
of the outer binary ($\beta_0$) are conserved. Also, whenever 
$\sqrt{\epsilon} \ll 2 \beta$, $a_1\epsilon \cos^2 I$ is a constant. 

The secular evolution is a result of coherent additions of
perturbations from the tertiary to the inner binary. Therefore, any mechanism that
perturbs the phase relation of the system could modify the evolution
of $\epsilon$ and $g_1$. For black hole systems,
an important effect comes from the general relativistic periastron
precession of the inner binary. Inclusion of this effect can be found
in Miller \& Hamilton (2002) and Blaes, Lee \& Socrates (2002). In
summary, contribution of the periastron precession in the
first-order post-Newtonian approximation can be added to the doubly-averaged Hamiltonian as
$\it {H}_{\rm PN}=-k\theta_{\rm PN}/\sqrt{\epsilon}$, where,
\begin{equation}
\theta_{\rm PN}=8\times
10^{-8}\frac{(M_1/\mbox{\msun})^2}{m_2/\mbox{\msun}}\left(\frac{b_2}{a_1}\right )^3\frac{1}{a_1/{\rm
AU}},
\label{tPN}
\end{equation}
and $k=3Gm_0m_1m_2a^2_1/(8M_1b^3_2)$ is a quantity related to the
evolution time scale.  Apparently the influence of the periastron precession
(abbreviated as PN effect) is the largest (in the sense that
$\theta_{\rm PN}$ is the largest)  for 
systems consisting of a tight and massive inner binary and a light,
far-out third component. With the addition of this PN effect, the total Hamiltonian remains conserved. It  can be
 written as $H=kW$ (\cite{miller02a}), where  
\begin{equation}
W(\epsilon,g_1)=-2\epsilon+\epsilon\cos^2I+5(1-\epsilon)\sin^2g_1(\cos^2I-1)+\theta_{\rm
PN}/\epsilon^{1/2},
\label{W}
\end{equation} 
is a conserved quantity.
\subsection{Eccentricity Evolution in Absence of the GR Effect}

The trajectory of $e_1$ and $g_1$ in the phase space are known to be
determined by the values of $\alpha$ and $\beta$ for given initial
system parameters (\cite{lidov76}).   In the general case of initial
$e_1 \ne 0$,$1$ and in the absence of the GR effect, the quantities $e_1$,
$g_1$ undergo cyclic oscillations.  
A necessary condition
for a large change in $e_1$ is that  $I_0 > \cos^{-1}\sqrt{3/5} \sim
39^o $ for a restricted case where $m_0 \ll m_2\ll m_1$. This
condition still holds for arbitrary masses when the PN effect is
weak (as can be proved using eq.~[\ref{epsi_min}]).  The time scale for the system to swing from $e_1 \sim 0$ to
$e_1 \sim 1$ is estimated to be  
(\cite{innanen97}),
\begin{equation}
\tau_{\rm{evol}} \approx 0.16 f
\left (\frac{M_1}{m_2} \right )^{1/2}\left (\frac{a_2}{a_1}\right )^{3/2}\frac{(a_2/{\rm AU})^{3/2}}{(m_2/\mbox{\msun})^{1/2}}(1-e^2_2)^{3/2}\ \rm{yr},
\label{tau_evol}
\end{equation}
where $f \sim 0.42\log(1/{e_1}_0)/\sqrt{(\sin^2(I_0)-0.4}$ is
a quantity of magnitude a few, ${e_1}_0$ is the initial value of $e_1$, and
$I_0$ is the initial value of $I$. This time scale should be  much
longer than the orbital 
periods of both binaries because it is an orbital averaged effect.

The minimum value of
$\epsilon$ (denoted $\epsilon_{\rm min}$)  for  arbitrary masses $m_0$,
$m_1$, $m_2$, and arbitrary $\epsilon_0$ can be solved for using the fact that the
Hamiltonian is conserved, that is, $W(\epsilon_0,g_0)=W(\epsilon_{\rm min},g_m)$
(eq.~[\ref{W}]).  We are interested the  cases  \epsimin\ $\sim 0$
and \epsimin$\ll \epsilon_0$. 
It follows that $g_m=\pi/2$ for $I_0 \ne I_c$ (\cite{lidov76}), 
$\epsilon^4_{\rm
min} \ll \epsilon^2_{\rm min}$, and  $\epsilon_{\rm min} \ll 2\beta$.  The assumption that $\epsilon_{\rm min} \ll 2\beta$ is
 valid as long as there exists a solution for the critical initial
value $\cos I_c =-\sqrt{\epsilon_0}/(2\beta)$.  Under these conditions, the $\cos^2 I$ term in
$W(\epsilon_{\rm min},g_m)$ can be rewritten in terms of $\cos^2 I_0$ and
$\epsilon_{\rm min}$.  An explicit approximation formula for $\epsilon_{\rm
min}$  is then obtained  by solving the resulting quadratic
equation as
\begin{equation}
 \epsilon^{1/2}_{\rm min} \approx \frac{1}{2\Omega}\left\{ \theta_{\rm
 PN}+\sqrt{\theta^2_{\rm PN}+20\Omega\epsilon_0\left [\cos
 I_0+\sqrt{\epsilon_0}/(2\beta)\right ]^2 }\right \}.
\label{epsi_min}
\end{equation}
By using equation~(\ref{aecosI}), we have
\begin{equation}
\epsilon^{1/2}_{\rm min} \approx \frac{1}{2\Omega}\left[\theta_{\rm
PN}+\sqrt{\theta^2_{\rm
PN}+5\frac{\Omega}{a_1}\frac{(\alpha^2_0-\beta^2_0)^2}{\beta^2_0}}\right ].
\label{epsi_min_2}
\end{equation}
Here
\begin{equation} 
\Omega=5-2\epsilon_0+\epsilon_0\cos^2I_0+\frac{\theta_{\rm
PN}}{\sqrt{\epsilon_0}}+4\epsilon_0(\cos
I_0+\frac{\sqrt{\epsilon_0}}{2\beta})^2+5(1-\epsilon_0)(\cos^2 I_0-1)\sin^2(g_0).
\label{omega}
\end{equation}
For the restricted
case of $m_0 \ll m_2\ll m_1$, we have $\beta \rightarrow \infty$, $I_c
\rightarrow 90^o$. We recover equation (8) of Miller \&
Hamilton (2002) using their assumptions that $\epsilon_0\approx 1$,
$\theta_{\rm PN} \ll 3$ and $5\cos ^2I_0 \ll 3$.  We then also recover the classical limit
$e_{\rm max}=\sqrt{1-\epsilon_{\rm min}}=\sqrt{1-(5/3)\cos ^2 I_0}$ 
(e.g., Innanen et al. 1997) by setting $\theta_{\rm PN}=0$.

It follows from equation~(\ref{epsi_min}) that a condition for achieving
extremely small  \epsimin\ is  $\theta_{\rm PN} \ll 1$ and $I_0
\approx I_c >90^o$, which 
corresponds to $\alpha \approx \beta$ (eq.[\ref{aecosI}]).  Note that
$I_c >90^o$ means that the outer binary should
initially be in a retrograde orbit with the inner one. This is
consistent with what we obtained in the classical limit ($\theta_{\rm
PN}=0$), where a system can evolve to $e=1$
 starting from any initial eccentricity if
$\alpha = \beta$ (or $I_0=I_c$) (see details in \cite{lidov76}).  The critical values of $I_0$ above which the system can have a significant
change in $\epsilon$ can also be derived.  If we demand $\epsilon_{\rm
min} \le 0.5 \epsilon_0$,   we would obtain from equation~(\ref{epsi_min}) $\cos^2 I_0 \le 3/5$, or $I_0 \ge 39^o$ for $\epsilon_0 \sim
1$ and $\theta_{\rm PN} \sim 0$. This is the same as the known classical limit for the restricted problem.

Two important constants can be
 derived from equation~(\ref{epsi_min}) for triple systems with
 arbitrary masses but negligible PN effect.  As $\theta_{\rm PN}
 \rightarrow 0$, $\Omega$ is a constant  $\sim 3$  for $\epsilon_0
 \approx 1$ and $I_0 \sim I_c$,
\begin{equation}
a_1\epsilon_{\rm min} \approx \frac{5}{12}\frac{(\alpha^2_0-\beta^2_0)^2}{\beta^2_0}.
\label{aepsm}
\end{equation}
It thus follows from equation~(\ref{aecosI}) that the mutual
inclination 
angle at \epsimin\  (denoted $I_m$) satisfies 
\begin{equation}
\cos^2I_m \approx 3/5,
\label{cosIm}
\end{equation}
 for $\sqrt{\epsilon_{\rm min}} \ll 2\beta$.

\section{Influence  of Gravitational Radiation Reaction (GR Effect) }
\label{GRR}
In this section, we discuss how the  GR effect damps the Kozai
oscillations and eventually leads to the merger of the system. As the
gravitational radiation frequency depends directly on the quantity
$a_1\epsilon$ ($\gamma^2$, or the  semi-latus rectum) (section~\ref{eccn}, eq.~[\ref{f_GW}]),  we focus our discussions on the evolution of this
quantity.   We first discuss the equations that govern the evolution
of the system where the GR effect is included.  We then derive the
minimal $\epsilon$ (\epsimin) obtained at the first Kozai cycle and then determine the evolution
of \epsimin\ during subsequent cycles, depending on the roles  of the PN effect and the GR
effect.  The evolution of the frequency of the gravitational wave
emitted by the inner binary can then be determined.

\subsection{Equations of Motion}
\label{GP}
The presence of gravitational radiation  has important impacts on the eccentricity
evolution of the inner binary.  First, the
radiation carries away energy and angular momentum and tends to shrink and circularize the orbit.  The
decay of the orbit subsequently changes the values of $\alpha$
and $\beta$ and thus modifies 
the secular evolution trajectory of the system. In particular,  the effect of periastron precession becomes stronger
rapidly with the decay of the orbit (see eq.~[\ref{tPN}]).  This
results in rapid oscillations of $g_1$ that could destroy the secular
oscillation of $\epsilon$.  As for the outer binary, the orbit remains
 much wider than the inner one, so its energy (equivalently $a_2$)
and magnitude of
angular momentum ($\beta_0$, or equivalently $e_2$) can therefore still be treated as 
conserved  quantities.

The GR effect is the strongest at $\epsilon \sim
\epsilon_{\rm min}$ within each Kozai cycle because of the strong
dependence of the merger time scale $\tau_{\rm GR}$  on $\epsilon$
 (eq.~[\ref{tau_GR}]).  When the PN effect is negligible and the GR
effect does not affect the Kozai cycles dramatically, we have $a_1 \propto \epsilon^{-1}_{\rm min}$ according to
equation~(\ref{aepsm}).  The evolution of the parameters $\theta_{PN}$, $\tau_{\rm evol}$, and
$\tau_{\rm GR}$ with the decay of the orbit  can then be  summarized as follows. 
\begin{eqnarray}
\theta_{\rm PN} \propto a^{-4}_1 \propto \epsilon^4_{\rm min}\\
\tau_{\rm evol} \propto a^{-3/2}_1 \propto \epsilon^{3/2}_{\rm min}\\
\tau_{GR}\propto a^4_1 \epsilon^{7/2}_{\rm min} \propto \epsilon^{-0.5}_{\rm min}.
\label{param_evol_equ}
\end{eqnarray}
It is apparent that, as the orbit of the inner binary  shrinks and circularizes,
the value of $\tau_{GR}$ decreases proportional to 
$\epsilon^{-0.5}_{\rm min}$ while the time the system spends  at
\epsimin\ ($\tau^{'}_{\rm evol}=\tau_{\rm evol}\sqrt{\epsilon_{\rm min}}$) increases
proportional to $\sim  \epsilon_{\rm min} ^{2} $.  When $\tau_{\rm GR}$ 
becomes of order $\tau_{\rm evol}\sqrt{\epsilon_{\rm min}}$, 
the gravitational merger will take place within one Kozai cycle near $\epsilon \sim\
$\epsimin.

The evolution of the system can be calculated through a set of hybrid
equations  that combines the equations of
motion derived from the conserved Hamiltonian $W(\epsilon,g_1)$ 
  (Lidov 1976, eq.~[32]--[33]) including the PN effect and that from
quadrupole  gravitational
radiation (\cite{peters64}, eq.~[5.6]--[5.7]). These are the same
equations used in Blaes, Lee, \& Socrates (2002).  
The evolution equations govern the four parameters $\epsilon$, $g_1$,
$a_1$,
and $\alpha_0$ and are given by
\begin{eqnarray}
\frac{d\epsilon}{dt} & =&  -10\kappa_E a^{1.5}_1\sqrt{\epsilon}(1-\epsilon)(1-\cos^2I)\sin(2g_1)+\frac{\kappa_G}{a_1^4}\frac{1-\epsilon}{\epsilon^{2.5}}(\frac{425}{304}-\frac{121}{304}\epsilon)\label{de_dt},\\
\frac{dg_1}{dt} & = & \kappa_Ea^{1.5}_1\left\{\frac{1}{\sqrt{\epsilon}}\left [4\cos^2I+(5\cos 2g_1-1)(\epsilon-\cos^2I)\right ] \right . \nonumber \\ & & \mbox{} \left.  +\frac{\cos I}{\beta}\left [2+(1-\epsilon)(3-5\cos 2g_1)\right ]+\frac{\theta_{\rm  PN}}{\epsilon}\right \}\label{dg_dt},  \\
\frac{da_1}{dt}& =& -\frac{6}{19}\frac{\kappa_G }{a^{3}_1}\frac{1}{\epsilon^{3.5}} (\frac{425}{96}-\frac{61}{16}\epsilon+\frac{37}{96}\epsilon^2)\label{da_dt},\\
\frac{d\alpha_0}{dt}& =&-\frac{3}{19}\frac{\kappa_G}{a^{3.5}_1}\frac{1}{\epsilon^2}(\frac{15}{8}-\frac{7}{8}
\epsilon) \frac{\sqrt{a_1\epsilon} +\beta_0\cos^2 I}{\alpha_0}\label{dH_dt},
\end{eqnarray}
where $\theta_{\rm PN}$ is defined in equation~(\ref{tPN}), and 
\begin{eqnarray}
\kappa_{\rm E}& =& 7.4554\times
10^{-8}(\frac{m_2}{M_1})^{0.5}\frac{(m_2/\mbox{\msun})^{0.5}}{(a_2/{\rm
AU})^3(1-e^2_2)^{1.5}}\frac{1}{{\rm
AU}^{1.5}}\label{kee},\\
\kappa_{\rm G}&=& 7.8218\times 10^{-26}\frac{m_0m_1M_1}{\mbox{\msun}^3}{\rm AU}^4\label{kg}.
\end{eqnarray} 
The evolution of the angular momentum $\gamma$   
can be written as (see eq.~[\ref{de_dt}],[\ref{da_dt}]),
\begin{equation}
\frac{d\gamma}{dt} =  -5\kappa_Ea^2_1\sin^2I\sin 2g_1
-\frac{3}{19}\kappa_Ga^{-3.5}_1\epsilon^{-2}(\frac{15}{8}-\frac{7}{8}\epsilon^2).
\label{dgam_dt}
\end{equation}

Equations~(\ref{de_dt})--(\ref{dH_dt}) provide a valid description of
 the evolution as long as  the energy ($a_1$) is approximately
conserved within each
cycle, and as the total
angular momentum vector of the triple system is conserved.  These are  necessary conditions
 for the validity of the equations
derived from the conserved Hamiltonian and the following considerations
 justify them for the systems that interest us.

We have assumed that the triple systems in
globular clusters consist  of stellar mass BHs with initial orbital
separations on the order of AU, and moderate initial $\epsilon \sim 1$
(see the 
discussion of parameters in section~\ref{eccn} below).  The systems are required to obtain extremely small
 \epsimin $\sim 0$  during their secular evolution  in
 order to merge well before the system is disrupted by
 interactions with field stars
(\cite{miller02a}). Equation~(\ref{de_dt}) and (\ref{dg_dt}) indicate 
that at each Kozai cycle, such systems would spend most of their time
at moderate $\epsilon \sim 1$, and a very small fraction of the time ($\propto
\sqrt{\epsilon_{\rm min}}$) at extremely small values (see also
Innanen, et al. 1997).  Note that gravitational radiation reaction is
negligible at moderate $\epsilon$ and $a_1$ as
$\kappa_G \ll \kappa_E$  (eq.~[\ref{kee}],[\ref{kg}]) for such
systems. The energy (equivalently $a_1$) can thus be treated approximately as conserved
for most of the cycles (see also eq.~[\ref{da_dt}]). 

The total angular momentum vector can also be treated as conserved within each
cycle as long as \epsimin$\sim 0$.  It is known that, for an isolated
system under gravitational radiation reaction, the  angular momentum
loss is negligible ($\propto \kappa_G\epsilon^{1.5}_{\rm min}$)
at \epsimin$\sim 0$
(eq.~[\ref{dgam_dt}], second term).  Moreover, gravitational radiation reaction  will not change the direction of the angular
momentum of the inner system.  It is thus conceivable that as
long as \epsimin$\sim 0$, the evolution of the angular
momentum of the inner binary ($\gamma$) is negligibly affected by the GR
effect, and the vector of the total
angular momentum can be treated as conserved throughout the
cycles.  For our systems, the GR effect starts to  dominate at
\epsimin$\sim 0$ because of the extremely small \epsimin\ required for
the system to merge before the system is disrupted by a field star. These justify the validity of equations~(\ref{de_dt})--(\ref{dH_dt}).

\subsection{Estimation of \epsimin}
\label{epsm}
The minimum value of $\epsilon$ that  a system can reach within
its first Kozai cycle  (denoted \epsimin)  is best estimated using
equation~(\ref{de_dt}).  At
$\epsilon=\epsilon_{\rm min}$,  $d\epsilon/dt=0$,  we obtain, 
\begin{equation}
\sin^2g_m=\frac{1+\sqrt{1-(\gamma_0\epsilon^{-3}_{\rm min}a^{-5.5}_1\sin^{-2}I_m)^2}}{2},
\label{gm}
\end{equation}
where 
\begin{equation}
\gamma_0=\frac{425}{3040}\frac{\kappa_G}{\kappa_E}=1.467\times
10^{-19}\frac{\mbox{\msun}^3}{m_0m_1M_1}\left (\frac{M_1}{m_2}\right )^{1/2}\frac{(a_2/{\rm
AU})^3(1-e^2_2)^{3/2}}{(m_2/\mbox{\msun})^{1/2}}{\rm AU}^{5.5},
\label{gam0}
\end{equation}
and where $\sin^2I_m$ is related to \epsimin\ through 
equation~(\ref{cosI}).  We estimate
$\epsilon_{\rm min}$ by assuming that $a_1$ is a constant and solving
implicitly the equation
\begin{equation}
\Delta W=W(g_0,\epsilon_0)-W(g_m,\epsilon_{\rm min})=0.
\label{Wmin}
\end{equation}
In our numerical work, we actually solve for \epsimin\  by minimizing $(\Delta
W)^2$ using a downhill simplex method developed by  Nelder and Mead (1965)
(available in Matlab).  This
method has proved to be robust in finding local minima.  However,
because of the existence of multiple solutions to equation~(\ref{Wmin})
(with at least one other solution $\epsilon=\epsilon_0$ and $g_1=g_0$), a
good initial guess is essential for a well-converged solution.

We made our initial guess \epsimin\  by considering the
following two limits for $g_m$ based on equation~(\ref{gm}).  In the
first 
limit,  
gravitational radiation reaction is weak, or $a^{11/2}_1\epsilon^3 \gg
\gamma_0$, we have $g_m \sim 90^o$.  Equation~(\ref{epsi_min}) can
then serve
as a very good
approximation to \epsimin.  In the second limit, 
gravitational radiation is very important,  $g_m$
could deviate significantly from $90^o$.  The extreme limit occurs 
around  $I_0 = I_c$ or 
$\alpha_0 = \beta_0$ (eq.~[\ref{aecosI}]).  As with
the critical case of $\alpha=\beta$ in the classical limit, a system starting with
$\epsilon_0 \sim 1$ will pass through  an
unstable stationary point $\epsilon \sim 1$ and $g_1 \sim
\sin^{-1}\sqrt{2/5}$ and move on towards a stationary (and stable in
the classical limit) point $\epsilon
\sim 0$, and $g_c=
\sin^{-1}\left [1/(2\beta)\sqrt{(8\beta^2-1)/5}\right ]$
(\cite{lidov76}, eq.
[46]). If it were not
for the GR effect, the system would reach $e_1=1$  at this stable
point. With the presence of the GR effect, the angle $g_1$ will  be
steered away from this 
point while $\epsilon$ is steered away from a true zero value (see
also Fig.~\ref{epsi_g1_2}).  However, we expect the value of
$g_m$ to be  very close to $g_c$ (at \epsimin$\sim 0$,
$dg/dt \propto \epsilon_{\rm min} \sim 0$) (eq.~[\ref{dg_dt}],
ignoring 
the PN effect).  In  this limit, the minimum
$\epsilon$ can thus  be estimated using  $g_m \sim
g_c$ and assuming constant $a_1$. Specifically, we again use the  minimization
program introduced previously to  solve equation
\begin{equation}
a^{5.5}_1\epsilon^3_{\rm min}=\gamma_0\frac{1}{\sin^2I_m}\frac{1}{\sin 2g_c},
\end{equation}
for \epsimin, where $\sin^2I_m$ is related to \epsimin\  by
equation~(\ref{cosI}). 

Without knowing the solution ahead of time, we
simply feed values of \epsimin\  estimated for these two limits
 as initial guesses to the minimization program and then select the
solutions of $(\Delta W)^2=0$ with minimum residues.  In
all cases, the residues of the minimization are monitored to ensure correct convergence. 

\subsection{PN effect}
\label{PN}
The importance of the PN
effect increases rapidly with the decay of the
orbit  as  $\theta_{\rm PN} \propto a^{-4}_1$.  This effect has a direct impact
 on the oscillatory behavior of $g_1$ (eq.~[\ref{dg_dt}]), through
which it can  affect the evolution of
$\epsilon$ and thus $a_1$\epsimin\  (eq.~[\ref{de_dt}]).  The importance of
the PN effect to the evolution of $\epsilon$ and $a_1$\epsimin, however, depends on the
role of the GR effect.

The PN effect can be  neglected if the GR effect dominates in the evolution of $\epsilon$ before the PN effect becomes important in the evolution
of $g_1$.   Once the GR effect dominates, the evolution of $\epsilon$ decouples from that of
$g_1$ and the Kozai cycle is terminated.  In the cases we are interested
in, this transition occurs near \epsimin$\sim
0$.  When both the GR and the PN effects
are not important, the  quantity $a_1$\epsimin\  is roughly  a constant (eq.~[\ref{aepsm}]). After the GR effect dominates, it evolves according to equation~(5.11)
of Peters (1964) and remains  a constant as long as \epsimin$\sim 0$.   The overall evolution of  $a_1$\epsimin\   can therefore be conveniently described using Peters' equation with $a_1$ and
\epsimin\  of the
first Kozai cycle (see section~\ref{epsm}) as the initial data.

The PN effect has a significant impact on the evolution of $\epsilon$
 and $a_1$\epsimin\ 
 if it  dominates in the evolution of $g_1$ 
 before the GR effect becomes important.  In this case, a dominating
 PN effect will 
 introduce a fast oscillation in $g_1$ and thus in  $\epsilon$
(eq.~[\ref{dg_dt}]) and will terminate the Kozai cycle. The GR effect will
 eventually dominate in the evolution of $\epsilon$ and Peters' 
 formula can again describe the evolution of $a_1$\epsimin.  However,
 when the PN effect is important, the quantity
 $a_1$\epsimin\ can no longer be
 approximated as a
 constant before the GR
 effect dominates.  It will evolve to a larger
 value than that in the first Kozai cycle (see eq.~[\ref{epsm}]).    The results based on
Peters' formula with initial values  taken at the first Kozai cycle can therefore be used to set the lower
 bound on $a_1\epsilon_{\rm min}$.

An upper limit for the quantity $a_1$\epsimin\  when the PN effect is
important can be set based on
the following considerations. First,
the fast
oscillation induced by the PN effect will be damped away under the GR effect. In light of
equation~(\ref{epsi_min_2}), the quantity  $\Omega$ will evolve from
a rough constant of 3--5 towards $\Omega \sim \theta_{\rm
PN}/\sqrt{\epsilon_0} \rightarrow \infty$ as $\theta_{\rm PN}
\rightarrow \infty$ with the decay of the orbit.  As a result, \epsimin\ $\rightarrow
\epsilon_0$, that is,  the oscillation will be quenched. The time scale for the system under the PN effect to swing from
$g_1\sim  0$ to $\pi/2$ with a small change in $\epsilon$ and $a_1$
is
\begin{equation}
\tau_{\rm PN} \sim \epsilon /(\kappa_E \theta_{\rm PN} a^{1.5})\sim
5\times 10^{6}  \left (\frac{a_1}{\rm AU}\right )^{2.5}
\frac{\epsilon}{(M_1/\mbox{\msun})^{1.5}}\  {\rm yr}.
\label{tau_PN}
\end{equation}
Secondly,  neglecting the GR effect, the evolutions of $I$, and $g_1$  under a dominating PN effect can be written as
\begin{eqnarray}
\frac{dI}{dt} & = & -5\kappa_Ea^2\sin I\cos I(a_1\epsilon)^{-1/2}\sin 2g_1 \label{dIdt},\label{dI_dt}\\
\frac{dg_1}{dt} & = & \kappa_Ea^{1.5}_1\frac{\theta_{\rm  PN}}{\epsilon}.
\end{eqnarray}
where we have applied equation~(\ref{cosI}) and (\ref{dgam_dt}) to obtain equation~(\ref{dI_dt}).
It is apparent that the oscillation of $I$ with $g_1$ is also damped
rapidly as  $\theta_{\rm  PN}/\sqrt{\epsilon} \rightarrow \infty$ with
the decay of the orbit. Integrating over $g_1=[0,\pi/2]$ for $dI/dg_1$, we obtained
an estimation for the oscillation amplitude
\begin{equation}
\delta I \approx 2.5 \frac{\sqrt{\epsilon}}{\theta_{\rm PN}} \sin 2I.
\end{equation} 
An upper limit of $\delta I$ can be estimated by the fact that  
  $\theta_{\rm PN}/\sqrt{\epsilon} > 10\cos^2I_m \sim 6 $ as required for a dominating PN effect
  (eq.~[\ref{dg_dt}],[\ref{cosIm}]), and that $\sin 2I \le 1$.  Here, we
   have used the fact that the PN effect becomes dominant near \epsimin\ (or $I_m$) where $a_1$ decreases most significantly under the
  GR effect.   We thus estimate that under the dominating PN effect,
  we have $I_m <39^o+\max(\delta I) < 63 ^o$. An upper limit for the
  quantity $a_1\epsilon_{\rm min}$ can therefore be
  set according to equation~(\ref{aecosI}) 
\begin{equation}
a_1\epsilon_{\rm min} \le \frac{5}{4}\frac{(\alpha^2_0-\beta^2_0)^2}{\beta^2_0}.
\label{aemax}
\end{equation}
By comparing this limit with equation~(\ref{aepsm}), we simply demand
that the upper limit of  $a_1$\epsimin\  be 
three times its lower limit obtained with Peters' formula.

For each given set of initial system parameters, we consider the PN
 effect to be important  if $a_1\epsilon_{\rm
 min}$ obtained within the first Kozai cycle, which set the lower
 limit of this quantity during its evolution, satisfies 
\begin{equation}
(a_1\epsilon_{\rm min})^{37/14} \ge {\gamma_0} \left
(\frac{6}{\theta^0_{\rm PN}}\right )^{5/7},
\label{PND}
\end{equation}
where $\theta^0_{\rm PN}=\theta_{\rm PN} a^4_1$ is a constant.  This condition is obtained by the requirement that the GR effect is
not important for the evolution of $\epsilon$ (eq.~[\ref{de_dt}]) when
the PN effect starts to dominate the evolution of $g_1$ (eq.~[\ref{dg_dt}]), that is, 
\begin{eqnarray}
a^{11/2}_1\epsilon^3 \gg \gamma_0,\\
\label{ae_GR}
\theta_{\rm PN} \ge 6 \sqrt{\epsilon}.
\end{eqnarray}
Equation~(\ref{PND}) and the comparison of $\tau_{\rm PN}$ (eq.~[\ref{tau_PN}]) with
$\tau_{\rm GR}$ (eq.~[\ref{tau_GR}]) indicates that the PN effect is important only for systems
with sufficiently large $a_1$ and \epsimin.  As a result, the PN effect is relevant
only in low eccentricity cases in the LIGO band (cf. eq.~[\ref{f_GW}]).

\subsection{Evolution of Gravitational Wave Frequency }
\label{s_f_GW}
The frequency of gravitational waves ($f_{\rm GW}$)  emitted from an inspiraling
binary depends strongly on its orbital frequency $f_{\rm orb}$ and
eccentricity $e_1$,  where
\begin{equation}
f_{\rm orb}=\frac{\sqrt{GM_1}}{2\pi}a^{-1.5}_1.
\label{forb}
\end{equation}
In the limit of the quadrupole approximation, the
 power  at the $n$th harmonic of $f_{\rm orb}$ have been derived in
 equations~(19) and (20) of
Peters \& Mathews (1963) in terms of spherical Bessel functions. For a circular orbit, the power is concentrated
in the second harmonic.  For an eccentric orbit, the power spreads over a broad
frequency band and the maximum occurs at a harmonic $n_m\gg 1 $  for
 $\epsilon \sim 0$. For instance,
this  peak frequency could occur at the $n_m=10^6$ harmonic of the orbital
frequency for $\epsilon=2\times 10^{-4}$.  In this section, we discuss the evolution of the peak frequency
with the eccentricity $e_1$ (or $\epsilon$). 

We have derived an approximate expression for the ``peak
harmonic'' of order $n_m$ in terms of $\epsilon$ for convenience of our
calculation as follows. We first used
the formulae in Peters \& Mathews (1963) to calculate
$n_m$ for various values of $\epsilon =10^{-6}$--1 and we then applied a least
square fit to the data. An excellent fit to  $n_m(\epsilon)$
was obtained,
\begin{equation}
n_m = \frac{2(1+e_1)^{1.1954}}{(1-e^2_1)^{1.5}}.
\label{gne}
\end{equation}
This  formula preserves the correct limit of $n_m=2$ for the circular case of $e_1=0$. 
The relation of the peak frequency $f^m_{\rm
GW}$ to $\epsilon$ is obtained by combining equations~(\ref{forb}) and (\ref{gne}),
\begin{equation}
f^m_{\rm GW}(e_1)=\frac{\sqrt{GM_1}}{\pi} (1+e_1)^{1.1954}\frac{1}{(a_1\epsilon)^{1.5}}, 
\label{f_GW}
\end{equation} 
It indicates that the peak frequency $f^m_{\rm
GW}$ is inversely
proportional to the period of a circular Keplerian orbit with radius
equal to the actual orbit's 
semi-latus rectum $a_1\epsilon$.  

We actually are interested in $e_1$ as a function of $f^m_{\rm GW}$,
rather than $f^m_{\rm GW} (e_1)$, since the focus of this article is
the orbital eccentricity as the GW frequency evolves through the LIGO
band.  We compute  $e_1$  
from equation~(\ref{f_GW}),  by applying our knowledge of the evolution of
$a_1\epsilon$.  Let the initial system parameters be  $a_1={a_1}_0$,
$\epsilon =\epsilon_0$ ($e_1={e_1}_0$), and the minimum $\epsilon$
expected for the first
Kozai cycle be \epsiminn\   (eq.~[\ref{Wmin}])  and $e_m=\sqrt{1-\epsilon^0_{\rm min}}$.  We
consider  the following
three possibilities separately. (i) The frequency enters the  LIGO frequency band
before $\epsilon$ reaches  \epsiminn.  In this case, $a_1 \sim
{a_1}_0$, and $\epsilon$
lies in the range  $\epsilon_0$--\epsiminn.  (ii) The frequency  enters the
LIGO  band after \epsiminn\   was reached and the GR effects dominates
in the evolution of $\epsilon$  before
the PN effect becomes relevant.  The evolution of $a_1\epsilon$ is then the same as for an
isolated  system evolving  under gravitational radiation reaction with
initial values of $a_1={a_1}_0$ and
$\epsilon=\epsilon_{\rm min}$.  That is, (eq.~[5.11] in Peters 1964),
\begin{equation}
a_1{\epsilon}=({a_1}_0 \epsilon^0_{\rm min})\left (\frac{e_1}{e_m}\right )^{12/19}\left (\frac{1+121/304e^2_1}{1+121/304e^2_m}\right )^{870/2299}.
\label{aepsi}
\end{equation}
(iii) The frequency  enters the LIGO  band after \epsiminn\ was reached, but the PN effect
becomes important  before the GR effect dominates in the evolution of $\epsilon$. In this case, the lower
limit of $a_1{\epsilon}$ is obtained from  equation~(\ref{aepsi}) and the
upper limit is set to be three times this lower limit. The bounds for $e_1$
at a fixed $f_{\rm GW}$ can be  obtained by applying  these two
limits. 

\subsection{Numerical Examples}
\label{example}
\begin{figure}
\PSbox{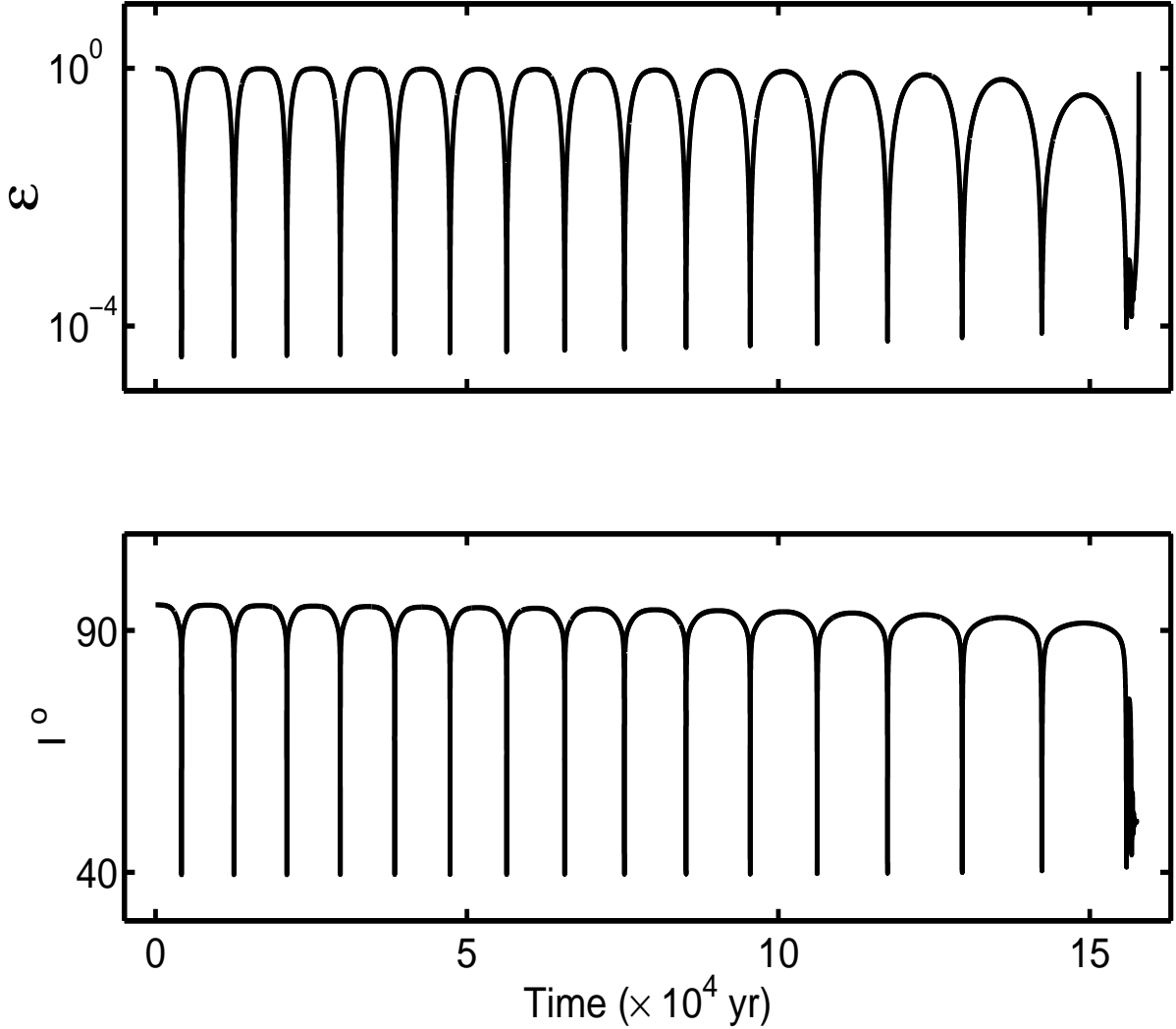 hoffset=-40 voffset=-160
hscale=90 vscale=100}{6.5in}{6.5in}
\caption{Secular evolution of $\epsilon=1-e^2_1$ of the inner binary and mutual
inclination angle $I$ 
between the inner and outer binaries. This evolution was computed
by integrating equations~(\ref{de_dt})--(\ref{dH_dt}) which include
contributions from Kozai mechanism, PN
periastron precession, and gravitational radiation reaction. The
initial system parameters are: $m_0=m_1=m_2=10$ \msun, $e_1=e_2=0.1$,
$a_1=10$ AU, $a_2/a_1=10$, $g_1=0$, and $I_0=95.3 ^o$. This is a
typical case in which the PN effect becomes important before the GR
effect dominates in the evolution of $\epsilon$, and that the system merges after many Kozai cycles.}
\label{e_t}
\end{figure}

\begin{figure}
\PSbox{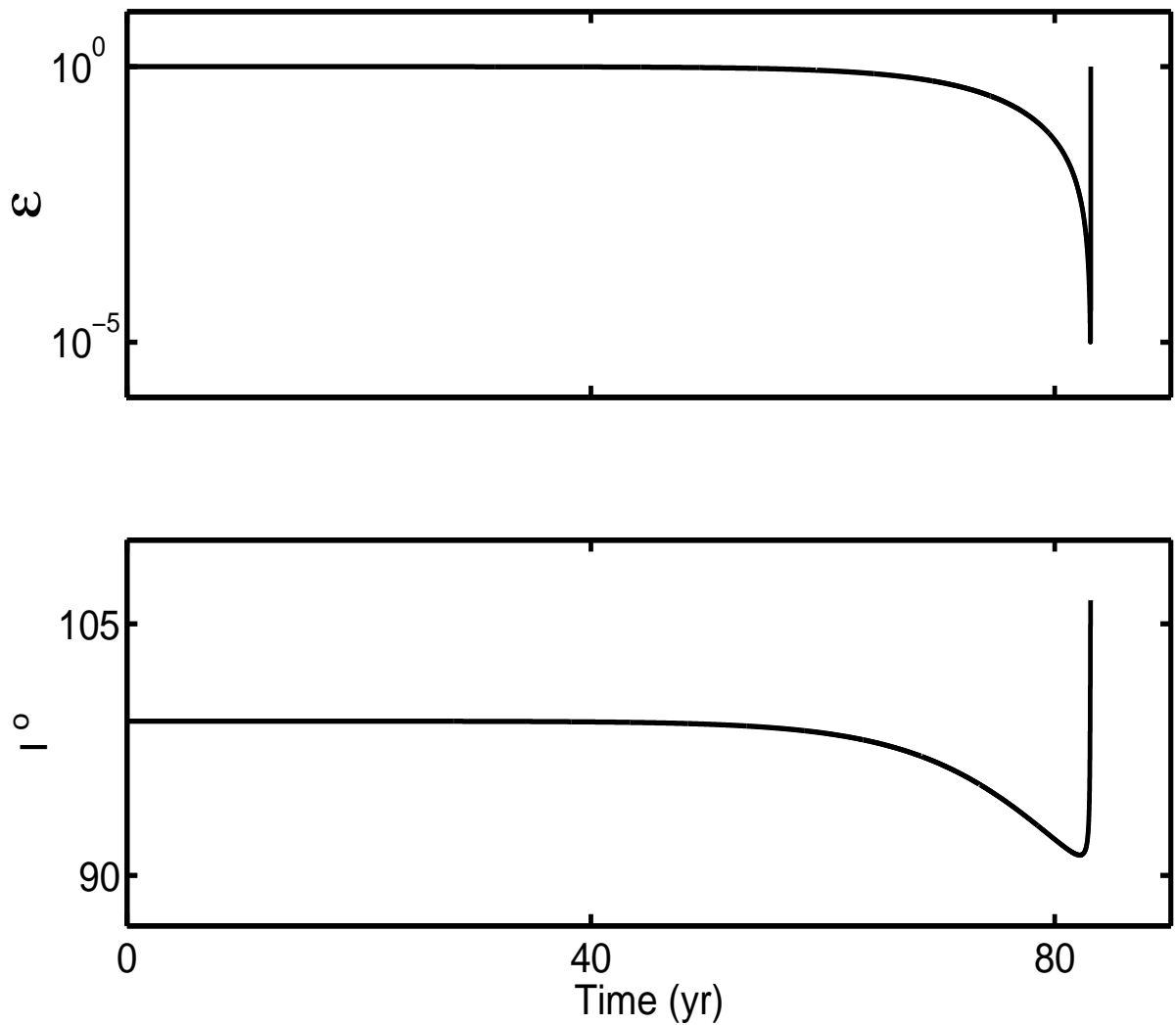 hoffset=-40 voffset=-160
hscale=90 vscale=100}{6.5in}{6.5in} 
\caption{Secular
evolution for  $\epsilon$ of the inner binary and of $I$, as computed
by integrating the evolution equations~(\ref{de_dt})--(\ref{dH_dt}) which include
contributions from the Kozai mechanism, PN
periastron precession, and gravitational radiation reaction. The
initial system parameters are: $m_0=m_1=m_2=5$ \msun, $e_1=0.01, e_2=0.51$,
$a_1=2.1920$ AU, $a_2/a_1=5$, $g_1=0$, and $I_0=99.2^o$. This is a
typical case where the GR effect dominates in the evolution within one Kozai cycle. }
\label{e_I_t_2}
\end{figure}

The numerical evolutions of $\epsilon$ and $I$ are shown for two
typical examples in Fig.~\ref{e_t} and
in Fig.~\ref{e_I_t_2}  for illustration.  These evolutions  were
obtained by integrating the  ordinary differentiation
equations~(\ref{de_dt})-(\ref{dH_dt}) for given initial data and then using equation~(\ref{cosI}) to derive the mutual inclination angle
$I$.  We used a  medium order Runge-Kutta method (available in Matlab as ode45)
for the integration.  

\begin{figure}
\PSbox{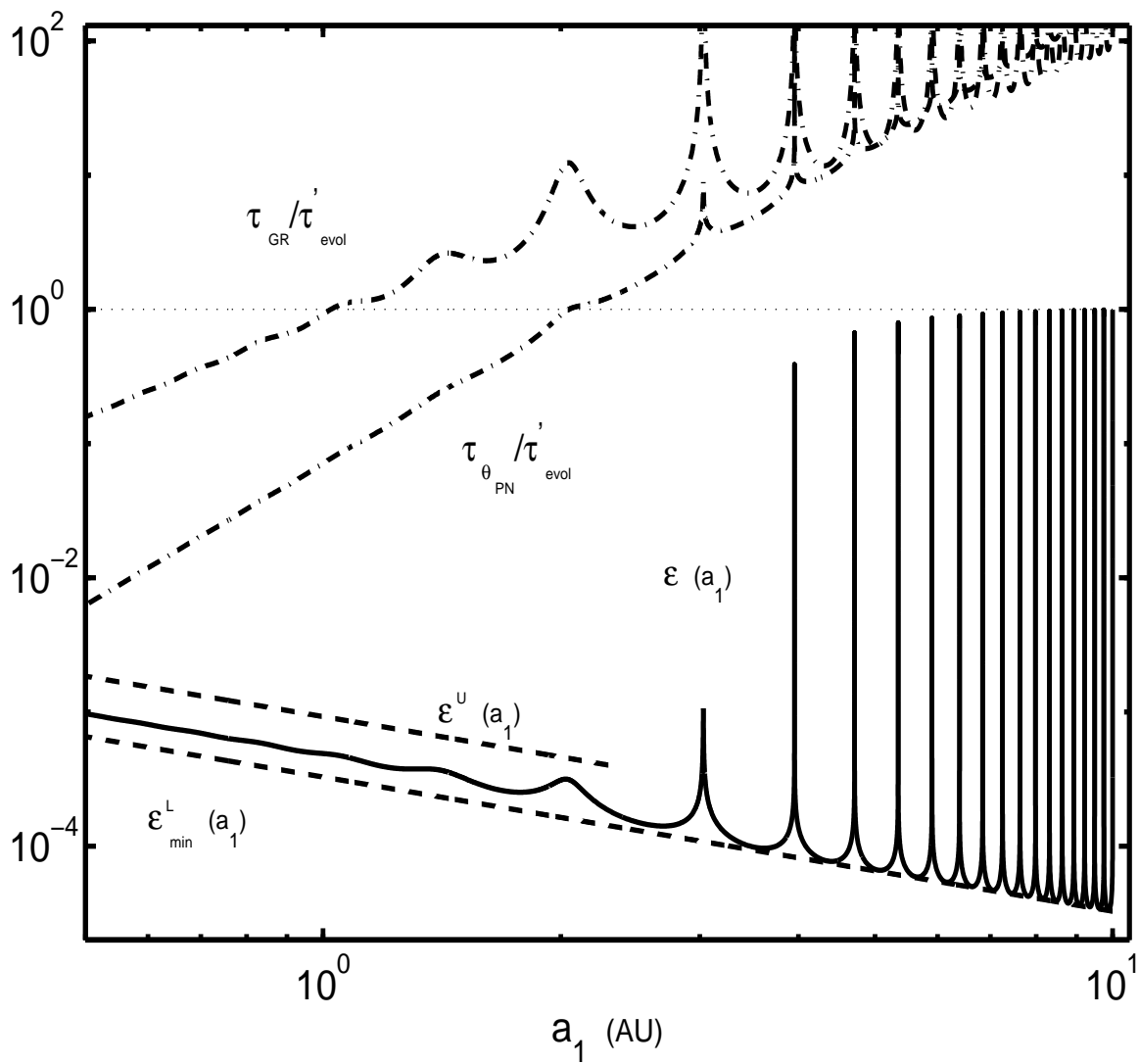 hoffset=-40 voffset=-160
hscale=90 vscale=100}{6.5in}{6.5in}
\caption{Evolution of
$\epsilon$ (solid line) and the time scales of gravitational merger
and the PN effect (dashed-dotted  lines) with the semimajor axis $a_1$
for the same 
initial parameters and evolution equations  as in Fig.~\ref{e_t}.  The evolution path of $\epsilon$
starts from the upper right of the solid line and moves towards the
left. The
upper and lower limits on $\epsilon_{\rm min} (a_1)$ based on equations~(\ref{aepsm}) and (\ref{aemax}) (dashed lines)  are also
shown.}
\label{param_evol}
\end{figure}

{\it The first example} (Fig.~\ref{e_t}) represents a case in which the system merges after many
Kozai cycles.  This is also a case where the PN effect becomes important  before the GR effect dominates in the evolution of $\epsilon$. The  system parameters are chosen to be $m_0=m_1=m_2=10$\msun, $e_1=e_2=0.1$,
$a_1=10$ AU, $a_2/a_1=10$, $g_1=0$, and $I_0=95.3 ^o$.  The gradual increase in the time scale of the Kozai cycle is
apparent. For most of the cycles,  $e_{\rm max} \sim 1$ (\epsimin\
$\sim 0$). Near the
\epsimin\  of the last Kozai
cycle,  fast damped oscillations of
$\epsilon$ (and $I$)  due to the PN
periastron precession are visible before a monotonically increase of
$\epsilon$ due to the dominating GR effect (see also Fig.~\ref{param_evol}).  The value of $I$
experiences damped oscillation after the PN effect becomes important and converges
to a constant after the GR  effect dominates in the evolution. Note also that before the last
cycle, the minimum $I$ is around $39^o$ as predicted by equation~(\ref{cosIm}).  

We show in Fig.~\ref{param_evol}, for the first example, the
evolution of $\epsilon$ with $a_1$ (solid line) and compare with the results based on the 
lower limit  of $a_1$\epsimin\ obtained with equation~(\ref{aepsi})
and a upper limit that is three times this lower limit as discussed
in section~\ref{PN} (dashed lines).    Also
shown in dashed-dot lines are the evolutions of the time scales of the
GR and the PN effects normalized
by the Kozai time scale $\tau^{'}_{\rm evol}$.  The evolution path of $\epsilon$
starts from the upper right of the solid line and moves towards the left.
It is apparent that $a_1$ remains essentially a constant within each
 cycle before $\tau_{\rm GR} /\tau^{'}_{\rm
evol} \sim 1$. The decay of $a_1$ occurs  mainly near
\epsimin\ of each cycle.  Damped oscillations are apparent near $\tau_{\rm PN}/\tau^{'}_{\rm evol} \sim 1$.  There is an excellent agreement
between our numerical results and the prediction in sections~\ref{epsm}-\ref{PN}. 

\begin{figure}
\PSbox{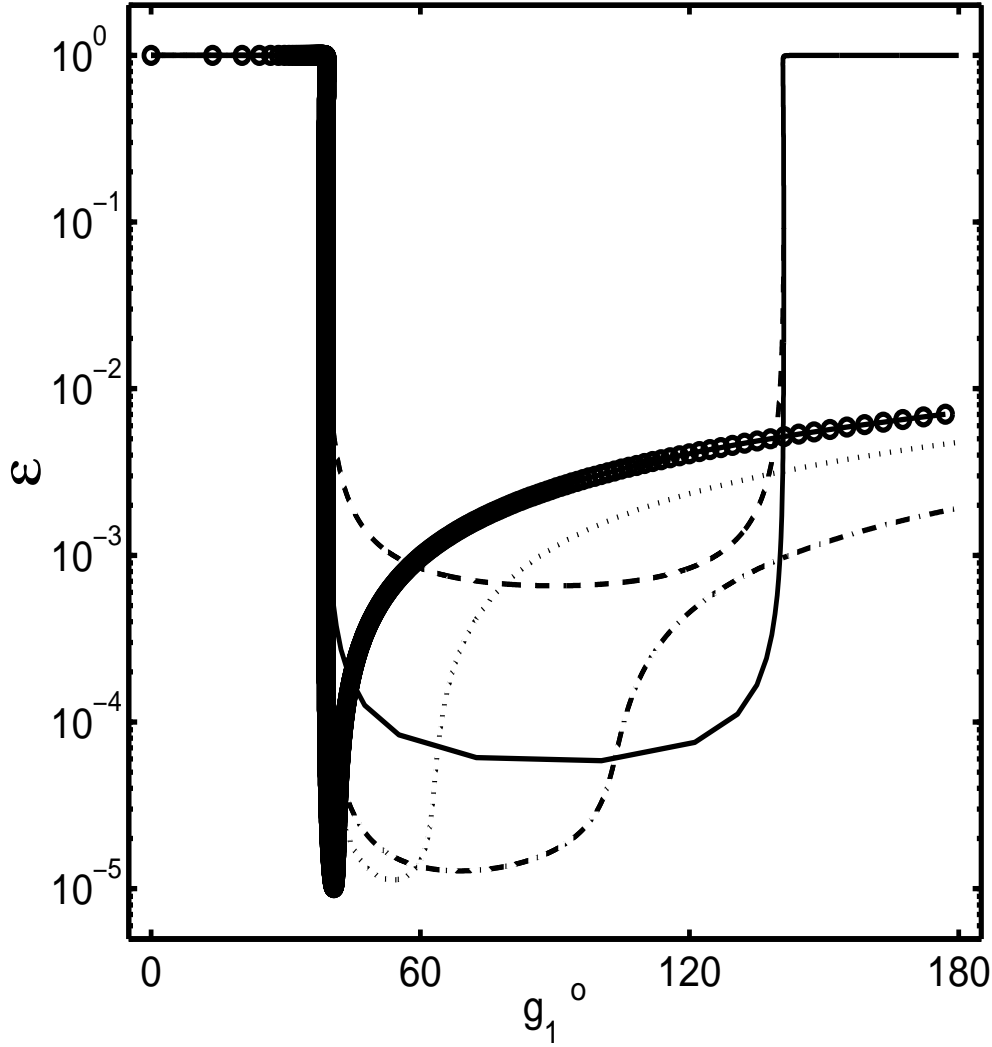 hoffset=-40 voffset=-160
hscale=90 vscale=100}{6.5in}{6.5in}
\caption{Phase diagram
of $\epsilon$ vs $g_1$ for the same initial parameters as in
Fig.~\ref{e_I_t_2} 
but with various $I_0$.  We have $I_0 =98^o$ (dashed line), $I_0=99.05^o$
(dotted line), $I_0=99.2^o$ (solid line with circles), $I_0=99.3^o$
(dash-dot line), $I_0=99.5^o$ (solid line).  For general cases,
\epsimin\ occurs at 
$g_1=g_m=90^o$. For $I_0 \sim I_c$, $g_m$ is significantly deviated from $90^o$. }
\label{epsi_g1_2}
\end{figure}

{\it The second example}  (Fig.~\ref{e_I_t_2})  represents a case where the
system merges within the first Kozai cycle.  The system parameters are chosen to be $m_0=m_1=m_2=5$\msun,
$e_1=0.01$, $e_2=0.51$, $a_1=2.1920$ AU, $a_2/a_1=5$, $g_1=0$, and $I_0=99.2
^o$. This is a case similar to the classical $\alpha=\beta$
case. That is, the system evolves along $\epsilon \approx 1$ until it
passes through an unstable stationary point at $g_1 \sim
\sin^{-1}\sqrt{2/5} \sim 39^o$  and then
evolves towards extremely small values of $\epsilon$ (Fig.~\ref{epsi_g1_2},
solid line with open circles).  The evolution is then quickly dominated by the gravitational radiation reaction near \epsimin.

\begin{figure}
\PSbox{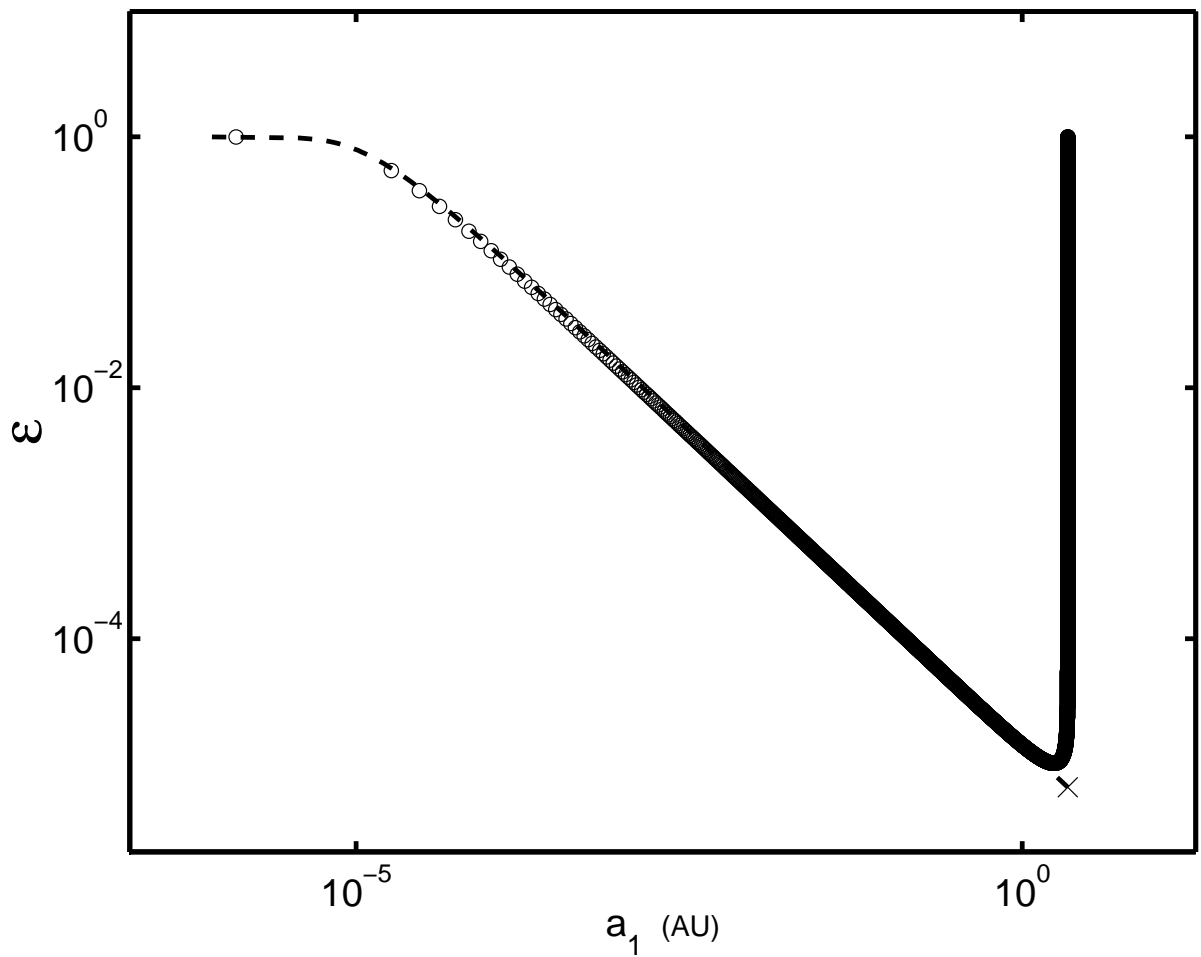 hoffset=-40 voffset=-100
hscale=90 vscale=90}{6.5in}{6.5in} 
\caption{Evolution of
$\epsilon$ with the semimajor axis $a_1$ (open circles) for the same 
initial parameters and evolution equations as in
Fig.~\ref{e_I_t_2}.  Note that the evolution path starts from the upper right
and moves towards the left.  The predicted \epsimin\ by equation~(\ref{Wmin}) at ${a_1}_0$ is shown by the 
symbol `X'.  The predicted evolution of $\epsilon$ vs $a_1$ from
equations~(\ref{aepsi}) is shown by the dashed line.  There is an
excellent agreement between the numerical values and the predictions. This
is a typical case where the GR effect dominates within one Kozai cycle
and the PN effect is negligible. } 
\label{a_epsi_2}
\end{figure}

The evolution of $\epsilon$ with $a_1$ (solid circles) for this second
example is shown
in Fig.~\ref{a_epsi_2}. The predicted \epsimin\ of the first Kozai
cycle calculated with the method
in section~\ref{epsm} at initial $
a_1={a_1}_0$ 
is shown in symbol `x'. The expected evolution of $\epsilon$ with $a_1$  by
equation~(\ref{aepsi}) starting from this predicted \epsimin\ and ${a_1}_0$ is shown by the
dashed line.  It is clear that there is an excellent agreement between
the predicted evolution of $a_1$\epsimin\
and the numerical values.  The same excellent agreement has been found for
all such cases we have investigated. 

The phase diagram for  the evolution path of $\epsilon$ vs $g_1$ for
systems with the parameters of our second example but with different
 initial values $I_0$ are shown  in
Fig.~\ref{epsi_g1_2}.   It is apparent that as
long as $I_0$ is away from the critical values $I_c$, the \epsimin\ is
reached around $g_1 \sim 90^o$. Near the critical value, $g_m \sim g_c$.

The evolutions of eccentricities $e_1$ as a function  of gravitational
wave frequencies $f^m_{\rm GW}$ for our first and second examples are
shown in Fig.~(\ref{e_f_1}) and
(\ref{e_f_2}).  In both cases, the GW frequency spans a wide range of eight orders of magnitude. This means that the GW's from these sources can be good
candidates for detections in the LISA band as well as LIGO.  For the case of many Kozai cycles in Fig.~\ref{e_f_1}, the eccentricity
drops sharply from $e_1 \sim 1$ at $f_{\rm GW} \sim 0.1$ Hz (the end
of the LISA band) where  the GR effect starts to dominate, and
is nearly zero at 10 Hz (the beginning of the LIGO band).  This is expected as equations~(\ref{f_GW})
and (\ref{aepsi}) predict that the eccentricities drop with the
frequency  roughly proportional to ${f^m}^{-19/18}_{\rm GW}$.  For
extreme cases as shown in Fig.~\ref{e_f_2}, the eccentricity remains
at the significant value of $\sim 0.9$ at $f^{m}_{\rm GW}=10$
Hz and could be extremely high at lower frequencies.

\section{Eccentricity Distribution}
\label{eccn}
The Kozai mechanism can drive the inner binaries of triple systems to
 extremely small $\epsilon$ and merge before disruption  by interactions with field
stars.  The time scale for disruption (the same as the stellar
 encounter time scale) is  given as (\cite{miller02a})
\begin{equation}
\tau_{\rm enc}\approx 6\times 10^5 n^{-1}_6 \frac{\rm AU}{a_2}\frac{10\mbox{\msun}}{M_2}\  \mbox{yr},
\label{tau_enc}
\end{equation}
where the number of stars in the globular cluster is $N=10^6n_6$. We assume $n_6=1$ in this paper.  Successful mergers of this sort therefore require 
\begin{eqnarray} 
\tau_{\rm evol} <\tau_{\rm enc}\label{tau_evol_enc},\label{t1}\\
\frac{\tau_{\rm GR} (a_1,\epsilon_{\rm min})}{\sqrt{\epsilon_{\rm
min}}} <\tau_{\rm enc}\label{tau_GR_enc}.
\label{t2}
\end{eqnarray}
Equation~(\ref{tau_evol_enc}) ensures that the system has completed at
least a half Kozai cycle to reach small
$\epsilon$. Equation~(\ref{tau_GR_enc}) ensures that the system can 
merge successfully before the next interruption.  The term on the
left-hand side of equation~(\ref{tau_GR_enc}) represents the total lifetime of the system which is  roughly  a factor of
$1/\sqrt{\epsilon_{\rm min}}$ the merger time scale $\tau_{\rm
GR}$. This is because the system only spends a significant fraction of
$\sqrt{\epsilon_{\rm min}}$ its time near \epsimin\ where the GR effect is
the strongest.

\subsection{Distribution of Initial \epsimin\  and $a_1$}
We have explored  the  parameter space of \epsimin\  and $a_1$ for
successful mergers driven by the Kozai mechanism in globular clusters
restricting ourselves to a 
triple system consisting of three 10 \msun\ BHs. We fix the initial parameters
$g_1=0$, and ${e_1}_0={e_2}_0=0.01$, and we have chosen $a_2/a_1=[20,10,5,3]$ and assume a
uniform distribution in the initial mutual inclination angles
$I_0$ and in the semimajor axis $a_1$.  The  minimum initial  value of 
$a_1$ was  set to $0.2$ AU so that during its interaction with  a third
body, the  recoil velocity associated with binary hardening is low enough
for the system to remain in the cluster (\cite{miller02a}). There was no presumed upper limit for $a_1$.  For each
given $a_1$, $I_0$, and $a_2/a_1$, we first calculated \epsimin\  with equation~(\ref{Wmin}) 
for each given $a_1$ and $I_0$. Equations~(\ref{tau_evol_enc}) and
 (\ref{tau_GR_enc}) were then evaluated to determine the permitted
parameter space. 

\begin{figure}
\PSbox{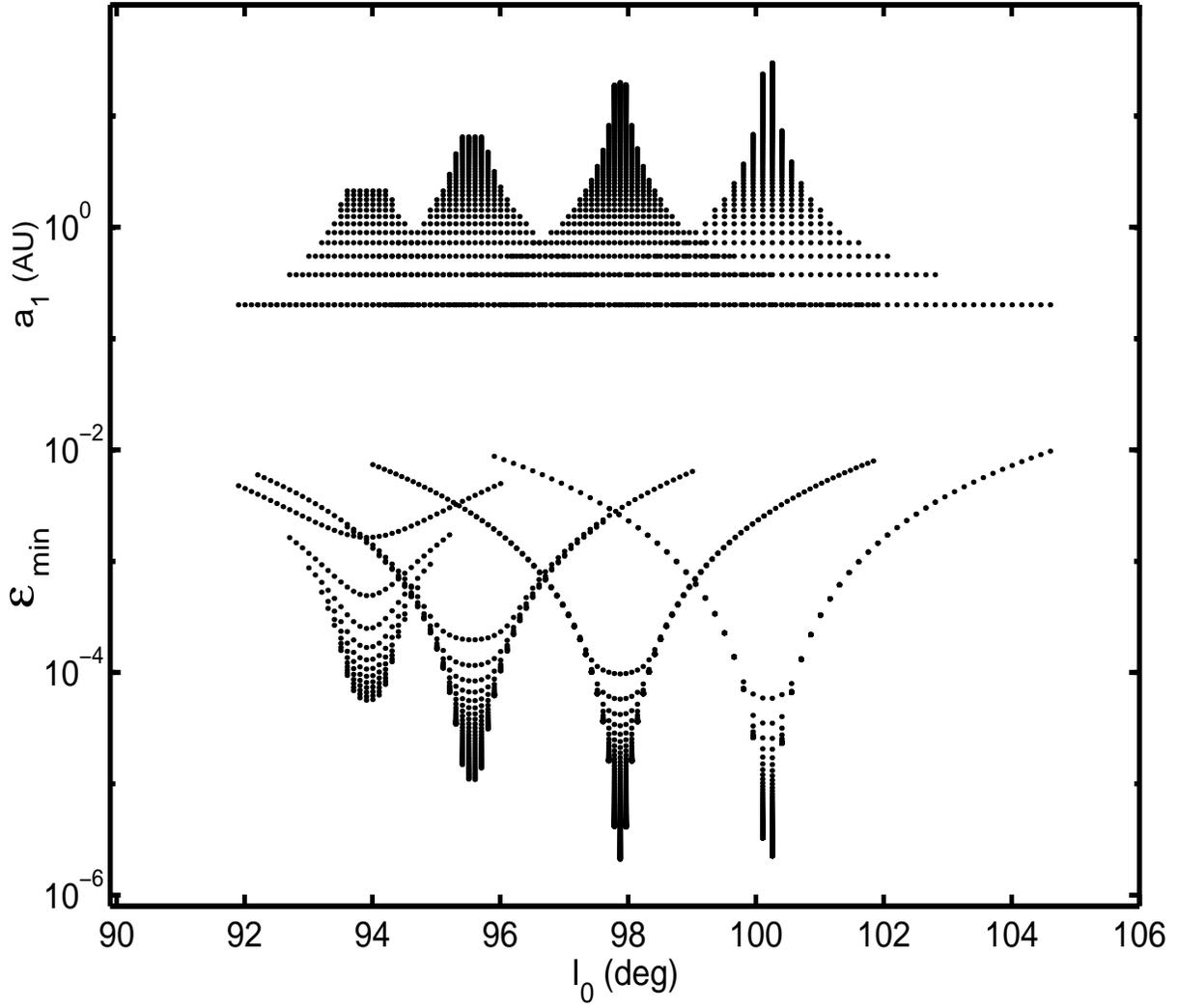 hoffset=-40 voffset=-160
hscale=90 vscale=100}{6.5in}{6.5in}
\caption{Parameter space for $a_1$  (upper set) and corresponding values of
\epsimin\ (lower set) vs $I_0$ for $a_2/a_1=20,10,5,3$ from left to
right (see text). }
\label{epsi_I}
\end{figure}

 We show in Fig.~\ref{epsi_I} the permitted values for $a_1$  (upper set) and
corresponding \epsimin\ (lower set) vs $I_0$.  
There are four sets of cone-shaped distributions, each corresponding to
a given $a_2/a_1$, with value decreasing  from the left to the
right. Each distribution centers on the critical value $I_0 \sim I_c$
which depends solely on $a_2/a_1$ (eq.~[\ref{Ic}]) as $\epsilon_0$ is
fixed.   There is a cut-off at a maximum $a_1$.   
This cut-off in $a_1$ is set by the constraint from
equation~(\ref{tau_evol_enc}).   For a fixed $a_2/a_1$,
larger $a_1$ implies larger $a_2$ and therefore  shorter $\tau_{\rm
enc}$ (eq.~[\ref{tau_enc}]). Beyond the maximum $a_1$, the time scale $\tau_{\rm
enc}$ becomes  so short that the
system is  disrupted before it can finish a half Kozai cycle to  reach \epsimin. For the same
reason, a larger
$a_2/a_1$ leads to smaller cut-off in $a_1$.

The overall cone shape in $a_1$
and \epsimin\  results from the fact that  the merger time scale should be less than the encounter time scale (eq.~[\ref{tau_GR_enc}]).   Near
the center of the cone where  $I_0 \sim
I_c$,  the system can reach  extremely small \epsimin\
(eq.~[\ref{epsi_min}]) which allows  maximum possible values of $a_1$
under  the constraint  of equation~(\ref{tau_GR_enc}).  The further $I_0$ is away from
the center ($I_c$), the larger  \epsimin\ is,  and so the smaller are
the $a_1$ values that  satisfy equation~(\ref{tau_GR_enc}).

The range of \epsimin\ within each distribution is caused by the PN effect.  In the classical limit (with no GR or PN effect), \epsimin\ depends only on the
values of $I_0$ and $a_2/a_1$.   It is independent of $a_1$
for each distribution, making 
\epsimin\  have a unique value at each $I_0$. This is
apparent for $\epsilon$ at $I_0$ far away from center of each
distribution for $a_2/a_1=3,5,10$ in Fig.~\ref{epsi_I}.
 A  significant contribution from the PN effect makes \epsimin\ depend
 on $a_1$, giving \epsimin\ a finite range (eq.~[\ref{tPN}],[\ref{epsi_min}]).  The PN effect is especially 
pronounced near the center of each cone ($I_0 \sim I_c$), where  the
contribution from the classical Hamiltonian is relatively smaller
(eq.~[\ref{epsi_min}]).  The finite range of \epsimin\  is more pronounced for
cones with larger $a_2/a_1$ values as the PN effect
is  more pronounced at larger $a_2/a_1$.  This also explains the
overall large \epsimin\ at larger $a_2/a_1$.

The \epsimin\ values estimated including the GR effect can
be several orders of magnitude  larger than those without the GR
effect in 
the region $I_0 \sim I_c$,  where the GR effect is the strongest.  This
makes little difference in finding the permitted parameter space for
$a_1$,  as
\epsimin\ will be extremely small in this region and
equations~(\ref{tau_GR_enc}) will be satisfied in either case.
However, the evolution of gravitational wave frequency depends
sensitively on our knowledge of $a_1$\epsimin. It is therefore necessary to include the GR effect.

\subsection{Distribution of Eccentricities in the LIGO band}
We proceed to investigate the distribution of eccentricities when the
emitted 
GW wave enters the LIGO band.  We restrict attention to stellar mass black holes in triple
systems and show only a representative case of triple systems consisting
of three 10 \msun\ black holes.   We consider a 
uniform distribution of initial  mutual inclination angle $I_0$,
 initial eccentricities of the inner
and outer binaries, ratios of the semimajor axis $a_2/a_1$, and
initial $g_1$. A summary of the parameter space we have investigated and the
representative values we have included can be found in Table~\ref{params}.

We choose the upper limit of $a_2/a_1$ to
be 30 based on equations~(\ref{tau_evol_enc}), as the
parameter range for $a_1$ diminishes with larger $a_2/a_1$ (see
Fig.~\ref{epsi_I}).  A general upper limit for the ratio
$a_2/a_1$ can also be set based on the  requirement that $\epsilon_0 \ge
\epsilon_{\rm min}$;  it follows from equation~(\ref{epsi_min}) that
\begin{equation}
\frac{a_2}{a_1} < 399 \left (\frac{a_1}{\rm AU} \right )^{1/3}\frac{\left
(m_2/\mbox{\msun}\right )^{1/3}}{(M_1/\mbox{\msun})^{2/3}}\frac{\epsilon^{1/6}_1}{\epsilon^{1/2}_2}.
\label{a2_a1}
\end{equation}
This expression is similar to 
the limit given by Blaes, Lee \& Socrates (2002) (eq.~[5])  with a
slight difference in the constant term and difference in the power of $\epsilon$ ( $1/6$  here
vs $1/2$ there). The lower limit for $a_2/a_1$ is set to be 3 based on the fact that
 the triple system is stable only if (\cite{mardling01}),
\begin{equation}
\frac{a_2}{a_1} > 2.8 \left [\left (1+\frac{m_2}{M_1}\right )\frac{1+e_2}{(1-e^2_2)^{1/2}}\right ]^{2/5}.
\end{equation}

We follow the same procedure described in the  previous subsection.  We first
calculate the permitted parameter space for $\epsilon_{\rm min}$ and
$a_1$.  The results (not shown) are many versions of the same
distributions of those shown in Fig.~\ref{epsi_I} for  various $a_2/a_1$. The
span of permissible $I_0$ is roughly 90--106$^o$, as required to obtain extremely small \epsimin. 
We then
calculate the expected eccentricity $e_1$ or its upper and lower limit 
at given frequency following the numerical 
procedures described in section~\ref{s_f_GW},  taking into account of
both the GR and PN effects. 

\begin{figure}
\PSbox{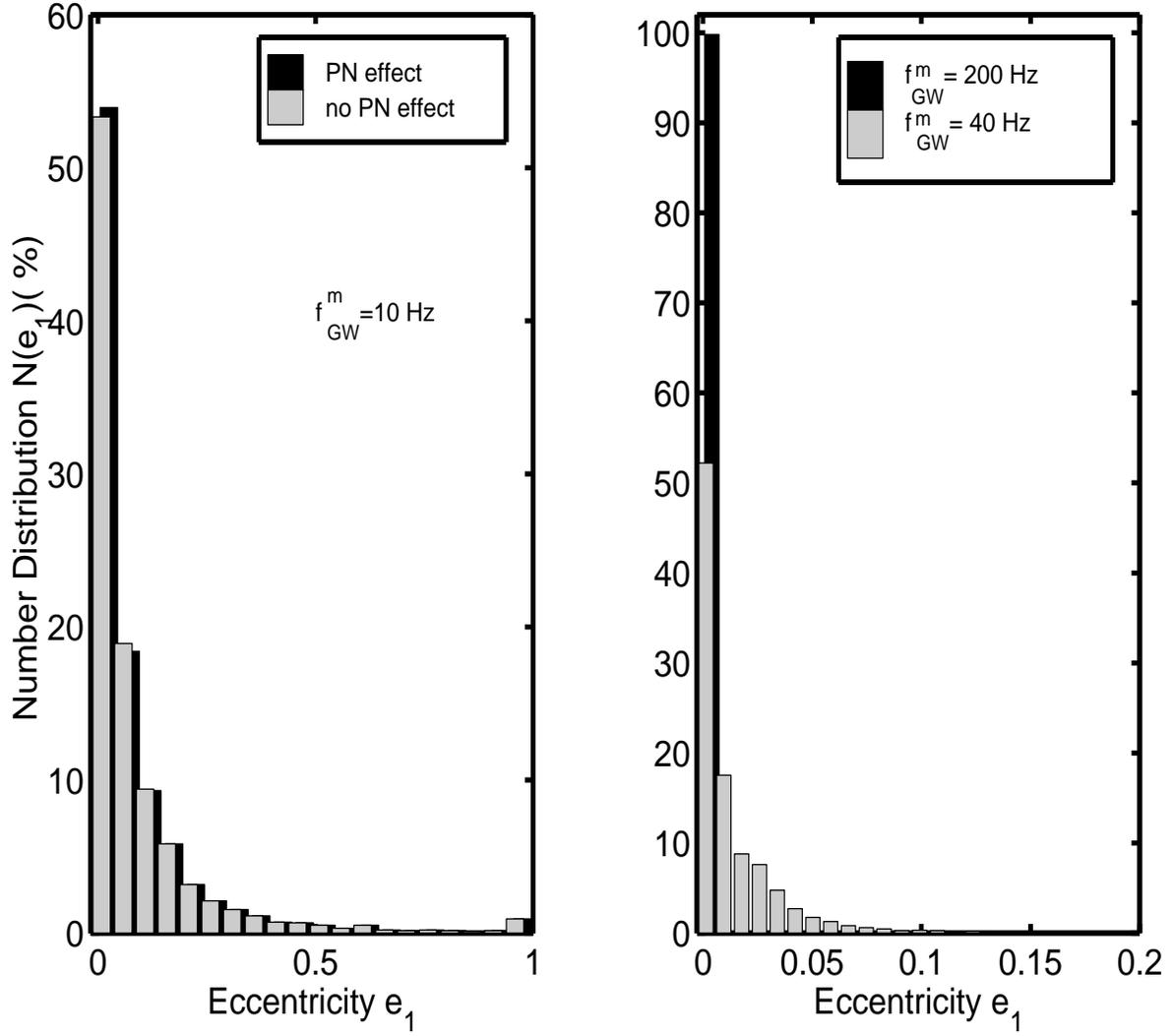 hoffset=-40 voffset=-160
hscale=90 vscale=100}{6.5in}{6.5in}
\caption{The distribution of eccentricities at $f^m_{\rm GW}=10$ Hz (left panel)
and 40 Hz, 200 Hz (right panel) for black hole mergers driven by the Kozai effect in a globular cluster. See
Table~\ref{params} for the parameter ranges used. Each parameter is
uniformly distributed in its range unless specified.}
\label{N_e}
\end{figure}

The number distribution of eccentricities at $f^m_{\rm GW} \sim 10$
Hz is shown as a histogram in the left panel of Fig.~\ref{N_e}.  The
distribution is shown in percentage relative to the total number of systems that would merge
before interruption by field stars.  The histogram in grey shade
represents the distribution without considering the impact of the PN effect on the evolution of
$a_1$\epsimin\ (see discussions in section~\ref{PN}).  The histogram
in dark shade includes the upper limit due to the  
impact of the PN effect. The x-axes of the histograms is shifted for a better
view.  It is apparent that the PN effect affects mostly the
low-eccentricity 
systems and it  makes  little difference in the histogram.  

At $f^m_{\rm GW}=10$ Hz,  around $70\%$ of the  merger systems have
eccentricity $<0.1$,  a little more than 50\% have eccentricities
$<0.05$.  About $\sim 2$ \%  have eccentricities $e_1 \sim
1$,  most of which are those 
with initial $I_0$ very close to the critical angle $I_c$,  as the
GW's  reach 
10 Hz before the
systems reach \epsimin\ within the first Kozai cycle.  These are the
type of systems studied in Fig.~\ref{e_I_t_2}, \ref{epsi_g1_2},
\ref{a_epsi_2}, and 
\ref{e_f_2}. 

We also show eccentricity distributions in the right panel of Fig.~\ref{N_e} for
$f^m_{\rm GW} =40$ Hz (grey shade) and $f^m_{\rm GW} =200$ Hz (dark shade).  At 40 Hz, all
eccentricities are well below 0.2. At 200 Hz, all are well below
0.02. This is not surprising as equations~(\ref{f_GW})
and (\ref{aepsi}) predict  that the eccentricities drop with the
frequency  roughly proportional to ${f^m}^{-19/18}_{\rm GW}$. This is consistent with the fact that the GR effect
circularizes the system in a very rapid fashion.

\section{Discussion}
\label{concl}
We have studied the evolution of eccentricities of black hole mergers
driven by the Kozai mechanism  in 
triple systems residing within globular clusters.  Eccentricity
distributions at gravitational wave frequencies relevant to LIGO
have been
presented. The evolution of these
systems were investigated including the  Kozai mechanism, the post-Newtonian periastron
precession effect, and gravitational radiation reaction.

\begin{figure}
\PSbox{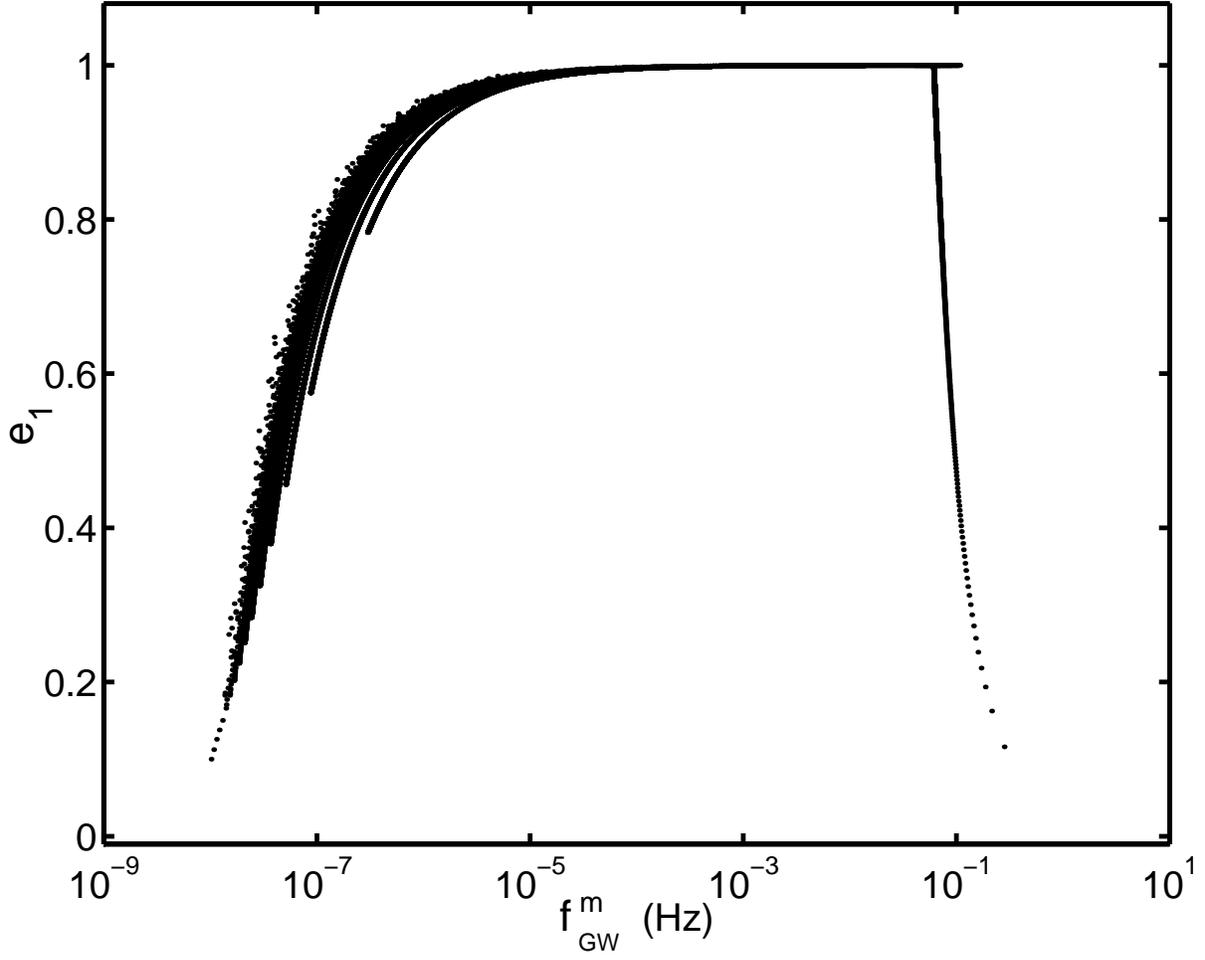 hoffset=-40 voffset=-100
hscale=90 vscale=90}{6.5in}{6.5in}
\caption{Evolution of
$e_1$ with gravitational wave frequency 
$f^m_{\rm GW}$ for the system shown in Fig.~\ref{e_t}. The $f^m_{\rm GW}$ spans
eight orders of magnitude. The eccentricity is nearly zero at and
above 10 Hz.} 
\label{e_f_1}
\end{figure}

\begin{figure}
\PSbox{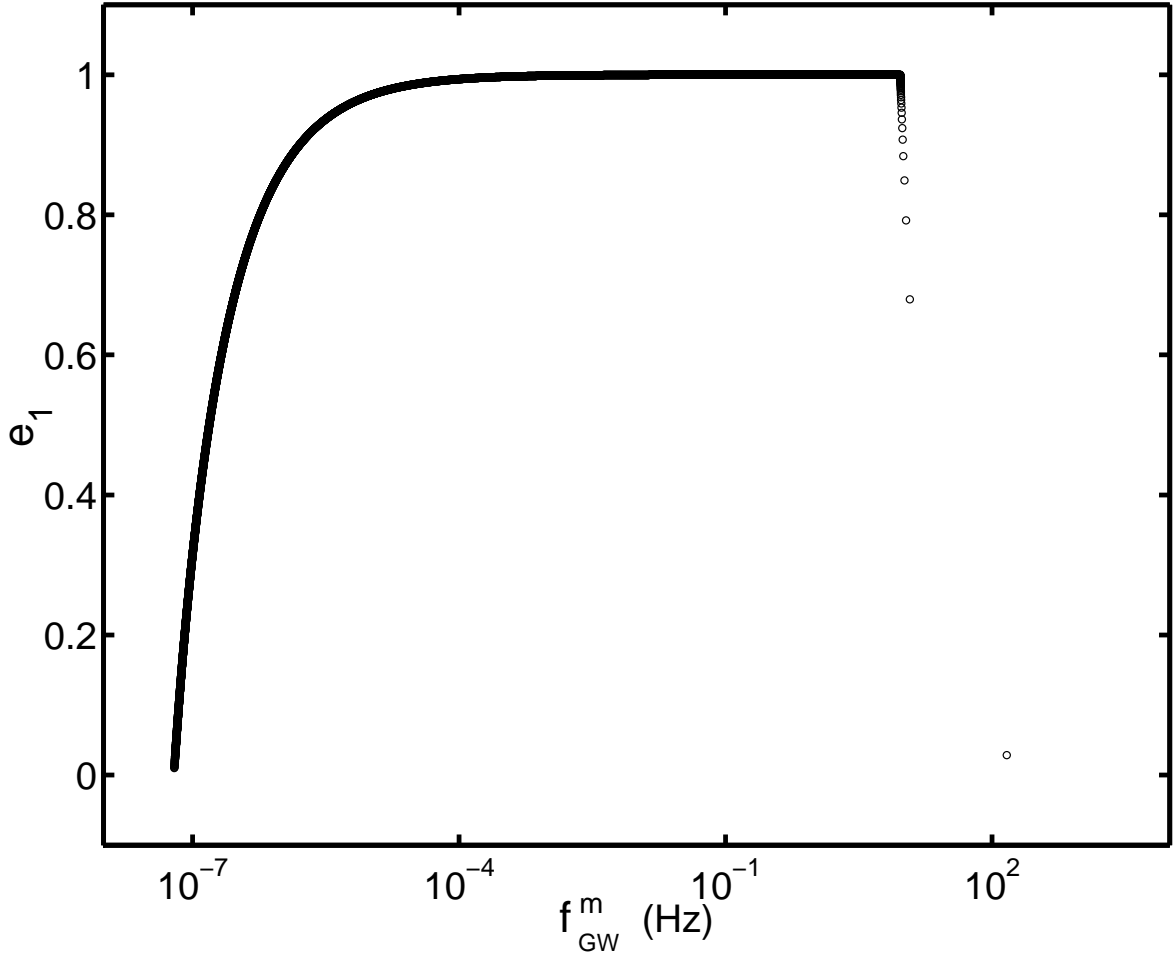 hoffset=-40 voffset=-100
hscale=90 vscale=90}{6.5in}{6.5in}
\caption{Evolution of
$e_1$ with 
$f^m_{\rm GW}$ for the system shown in Fig.~\ref{e_I_t_2}. The $f^m_{\rm GW}$
again spans
eight orders of magnitude. The eccentricity is around $0.9$ at 10 Hz.} 
\label{e_f_2}
\end{figure}

We
conclude that around 30\% of the systems possess eccentricities
$>0.1$ when the emitted gravitational waves  reach 10 Hz.  Around 2\% of
the systems are  extremely eccentric at 10 Hz.  However almost all our 
merger systems possess eccentricities well below 0.2 at 40 Hz, and
below 0.02 at 200 Hz (see also a brief discussion in Miller 2002).  These merger
systems, on the other hand, promise extremely high eccentricities at
the lower
frequencies of the LISA band (see Fig.~[\ref{e_f_1}],[\ref{e_f_2}]).   

About 30\% of our merger systems possess $e> 0.1$ at 10 Hz, the
lower end of the advanced LIGO frequency band (\cite{fritschel02}).  It is thus important to determine whether
it 
is possible to use LIGO's current circular-binary search templates  to
detect such systems.   For highly eccentric systems,
 the higher harmonics probably need to be taken into account when 
optimizing the search templates.  It is plausible that the eccentricities associated with
these systems are not important at 200 Hz or perhaps even at 40 Hz. It
is important however, to determine the limit of $e_1$ below which,
circular-binary templates need to be replaced by new, eccentric-binary
templates (see \cite{martel99}). 

The eccentricity distribution was calculated for triple systems
consisting of three 10 \msun\ black holes.  Similar  eccentricity
distribution, however, can be found for triple systems with individual masses in the range of 3--25
\msun\ known for the observed galactic stellar mass BHs
(\cite{bailyn98}).   We have assumed that, in globular clusters,  triple
systems with one or more BHs of mass $> 25$\msun\ are rare.  The mass parameters affect very little the values of \epsimin\ a system can reach (eq.~[\ref{epsi_min}],[\ref{tPN}],[\ref{gam0}]).  Its effect to the shape of
eccentricity 
distribution is also very weak 
compared with other parameters such as $a_1$, $a_2/a_1$, and
$\epsilon$ (eq.~[\ref{t1}], [\ref{t2}]). Any effects due to different masses will be further averaged out
if we include a uniform
distribution of masses within this range.

Our calculation of the gravitational radiation reaction is based on the Newtonian
quadrupole approximation. At 10 Hz, this approximation is still valid
 as $v/c \sim 0.1$ at the periastron. However, at higher frequencies, especially near 200 Hz in the LIGO frequency band, this
approximation breaks down as the speed  of the binary orbit is
approaching that of light.  However, all our binaries become so
circular well before they reach 200 Hz, that our results are probably still
relevant. 

\acknowledgments   
 We would like to thank Kip Thorne for an introduction to this project
and for his continuous encouragement and critical discussion on the
research.  We thank Tom Prince,  Albert Lazzarini, Barry Barish, 
for their support and discussion of this work.  We also thank Cole
 Miller, Maurice van Putten, Teviet Creighton, Alessandra Buonanno,
 and David Shoemaker  for very useful inputs.  Support for this
work was provided in part by NSF grant PHY-0071050 and by the LIGO
Laboratory under NSF cooperative
agreement PHY-0107417. This paper has been assigned the LIGO Document
 Control Center number LIGO-P020022-00-D.

\pagebreak


\clearpage
\newpage
\begin{deluxetable}{lcccccc}
\small
\tablecolumns{6}
\tablewidth{0pc}
\tablecaption{Parameter ranges investigated for Kozai-mechanism-driven BH mergers in globular cluster \label{params} }
\tablehead{
\colhead{ $M_{\rm BH}$ (\msun)} & \colhead{ $a_1$ (AU)} & \colhead{$a_2/a_1$} &
\colhead{$e_0$}  & \colhead{$g_1 (^o)$} &\colhead {$I_0 (^o)$}  &\colhead {$n_6$} }\tableheadfrac{} 
\startdata
10 & 0.2--30 & 3,5,10,20,30 & 0.01--0.901 & 0--90 &
85--110&  1 
\enddata 
\tablecomments {$M_{\rm BH}$ refers to black hole masses for all three
components;  $e_0$ refers to eccentricities for both inner and outer
binaries.  Numbers of steps for $a_1$, $e_0$,  $g_1$, and $I_0$ are 60,
4, 4, and 100 respectively. See the text for definitions. }
\end{deluxetable}

\end{document}